\documentclass[prd,twocolumn, superscriptaddress, nofootinbib, preprintnumbers]{revtex4-1}

\usepackage[bottom]{footmisc}
\usepackage{hyperref}
\usepackage[normalem]{ulem}
\usepackage[utf8]{inputenc}
\usepackage{amsmath}
\usepackage{graphicx}
\usepackage{braket}
\usepackage{color}

\newcommand{\ob}{\Omega_b}
\newcommand{\om}{\Omega_m}
\newcommand{\oq}{\Omega_Q}
\newcommand{\dq}{\delta_Q}
\newcommand{\uq}{u_Q}
\newcommand{\dm}{\delta_m}
\newcommand{\dmr}{\delta_m^{\mathcal{R}}}
\newcommand{\dmi}{\delta_m^{\mathcal{I}}}

\newcommand{\dl}{\delta_L}
\newcommand{\eps}{\epsilon}
\newcommand{\pd}{\partial}
\newcommand{\R}{\mathcal{R}}
\newcommand{\I}{\mathcal{I}}
\newcommand{\Lg}{\mathcal{L}}
\newcommand{\cq}{c_{{}^{Q}}^2}
\newcommand{\ca}{c_{a}^2}
\newcommand{\cs}{c_{s}^2}
\newcommand{\lb}{\left (}
\newcommand{\rb}{\right )}
\newcommand{\wq}{w_{{}_Q}}
\newcommand{\dmw}{\delta_{mW}}
\newcommand{\ks}{k_S}

\newcommand{\tmr}{T_m^\R}
\newcommand{\tmi}{T_m^\I}
\newcommand{\ter}{T_\eps^\R}
\newcommand{\tei}{T_\eps^\I}

\newcommand{\blr}{b^{L}_{\R}}
\newcommand{\bli}{b^{L}_{\I}}

\newcommand{\dsh}{{\hbox{-}}}
\newcommand{\lsim}{\mathrel{\rlap{\lower4pt\hbox{\hskip1pt$\sim$}}\raise1pt\hbox{$<$}}}
\newcommand{\gsim}{\mathrel{\rlap{\lower4pt\hbox{\hskip1pt$\sim$}}\raise1pt\hbox{$>$}}}

\begin{document}
	
\preprint{YITP-SB-18-43}
\title{Quintessential Isocurvature in Separate Universe Simulations}
\author{Drew Jamieson and Marilena Loverde \\{\it{\small  C.N. Yang Institute for Theoretical Physics, Department of Physics \& Astronomy, Stony Brook University, Stony Brook, NY 11794}}}

\begin{abstract}
	In a universe with quintessence isocurvature, or perturbations in dark energy that are independent from the usual curvature perturbations, structure formation is changed qualitatively. The existence of two independent fields, curvature and isocurvature, causes the growth rate of matter perturbations to depend on their initial conditions. The quintessence perturbations cause their growth to depend on scale. We perform the first separate universe simulations for this cosmology. We demonstrate that the power spectrum response and the halo bias depend on scale and initial conditions and that the presence of the isocurvature mode changes the mapping from these quantities to the halo auto- and cross-power spectra, and the squeezed-limit bispectrum. We compare the bias to several models, finding reasonable agreement with both a power-spectrum-response model with one free parameter and a model that fits two independent bias parameters for curvature and isocurvature sourced fluctuations. We also verify that simulation responses to pure isocurvature and pure curvature modes can be linearly combined to reproduce responses with different ratios of isocurvature and curvature. This allows our results to be used to predict the halo power spectrum and stochasticity with arbitrary large-scale curvature and isocurvature power spectra. In an appendix, we study the generation of quintessence isocurvature during inflation and show that a modified kinetic term is typically required to produce observable isocurvature modes in a field with $w_Q\approx -1$. 
\end{abstract}

\maketitle
	
\section{Introduction}
	
	The nonlinear regime of large-scale structure can provide a wealth of information about the initial conditions of our universe and its dynamical evolution. However, analytic calculations of nonlinear growth present formidable conceptual and technical challenges. The separate universe formalism has proven to be a useful tool for studying mode coupling in the evolution of large-scale structure, providing a conceptually simple framework that gives insight into the nonlinear regime. In the separate universe formalism, the nonlinearity of structure growth is studied through the effect that long-wavelength modes have on small-scale observables, such as the local power spectrum and local number densities of collapsed objects. Since the long-wavelength modes are far above the nonlinear scale, their evolution can be calculated using linear perturbation theory, while the small-scale observables, extracted from N-body simulations, can be determined deep into the nonlinear regime. 
	
	Separate universe techniques have been applied to studies of mode coupling in the production of primordial density perturbations during inflation \cite{Maldacena:2002vr,Creminelli:2004yq}, including isocurvature modes in multifield models \cite{Grin:2011tf}. The formalism was then developed for large-scale structure formation throughout matter and dark energy domination \cite{McDonald:2001fe,Sirko:2005uz,Wagner:2014aka, Dai:2015jaa, Hu:2016ssz, Li:2014sga, Chiang:2014oga, Manzotti:2014wca, Baldauf:2011bh, Baldauf:2015vio, Lazeyras:2015lgp, Gnedin:2011kj, Li:2014jra}. This approach was applied to investigations of halo assembly bias \cite{Paranjape:2016pbh},  adiabatic perturbations in the presence of clustering quintessence \cite{Chiang:2016vxa}, and the cosmological effects of massive neutrinos \cite{Chiang:2017vuk}. These latter two studies found that scale dependence in structure formation can arise from scale dependent growth of the long-wavelength matter perturbations. Although, the effect was relatively small in both cases.
	
	In this work, we apply these separate universe techniques to the case of clustering quintessence which sources isocurvature perturbations. Our goal is to study the clustering of matter on large scales in a broader context, in which the growth histories of long-wavelength matter perturbations can differ dramatically. For this purpose, we chose a model that makes the effects of history dependence large, allowing us to test our understanding of the scale dependence and time evolution of quantities such as the halo bias. The quintessence model is chosen as an academic example of a scenario with isocurvature perturbations that are important at late times and we do not require consistency of the model parameters with current data. We will, however, comment on the observational viability and physical imprint of this type of quintessence on the observed matter power spectrum, halo bias, and halo stochasticity.
	
	Quintessence isocurvature is an extension of standard $\Lambda \rm{CDM}$ cosmology, in which a scalar field is both responsible for the dark energy content of the universe, and also gives rise to primordial entropy perturbations. These perturbations are additional degrees of freedom for the initial configuration of energy densities and pressures in the early universe. Their presence affects the evolution of structure growth and halo formation in a scale dependent way. The effect can be made arbitrarily large or small by tuning the ratio of primordial curvature and isocurvature. 
	
	The effect of quintessence isocurvature on cosmic structure enables us to study the history dependence of structure formation. Matter perturbations that are sourced by different amplitudes of primordial curvature and isocurvature will evolve differently in time. Additionally, the clustering property of the quintessence introduces a new physical scale, the quintessence Jeans scale, which causes scale dependent evolution for different modes of the matter density perturbation. Quintessential scale dependent growth was first studied in \cite{Chiang:2016vxa}, using separate universe simulations with a model that contained purely adiabatic perturbations. In the current work we use the same model but with different parameters, allowing for isocurvature perturbations that are used to amplify the scale dependence in the evolution of the matter. This allows us to study structure formation in regions for which the large-scale matter and quintessence fields arrive at the same final state, but evolve through significantly different histories.
	
	This paper is organized as follows. Section~\ref{sec:isocurvature} presents the model of clustering quintessence, focusing on features of the model that allow for isocurvature perturbations that are conserved on superhorizon scales. Section~\ref{sec:SU} reviews the separate universe formalism and the numerical linear perturbation theory calculations that were used to fix the expansion histories of our separate universe simulations. Section~\ref{sec:sims} summarizes how the simulations were performed, the parameter choices that were made, and how the results were analyzed. Section~\ref{sec:Presponse} shows the results for the matter power spectrum responses, comparing them to 1-loop perturbation theory calculations. The halo biases measured from the simulations are presented in Section~\ref{sec:bias}. In Section~\ref{sec:bias} we also verify that linear combinations of the individual responses measured from simulations with pure isocurvature and pure adiabatic long wavelength modes reproduce the net response for simulations with different fractions of initial isocurvature perturbations. This allows us to generate predictions for halo bias and the squeezed-limit bispectrum for arbitrary initial curvature and isocurvature power and cross-power spectra without the need for additional simulations. Section~\ref{sec:biasmodels} compares the simulation bias results to models for scale-dependent bias.  In Section~\ref{sec:observables} we consider the long-wavelength power spectrum, clustering bias, and stochasticity arising from ensemble averaging over the long-wavelength primordial curvature and isocurvature modes. In the \hyperref[sec:generationmodes]{Appendix} we discuss the viability of some mechanisms to generate quintessence isocurvature perturbations in the early universe. 
	
\section{Isocurvature perturbations}
\label{sec:isocurvature}
	
	We work with a K-essence \cite{ArmendarizPicon:2000dh} type scalar field dark energy Lagrangian of the form,
	\begin{align}
		\label{eq:L}
		\Lg & = \frac{2\cq\Lambda}{1+ \cq} \lb \frac{X}{\Lambda} \rb ^  { \frac{1+\cq}{2 \cq } } - V\lb Q \rb \, ,  \\
		X & = -\frac{1}{2} g^{\mu\nu} \pd_\mu Q\pd_\nu Q \, . 
	\end{align}
	The quintessence field $Q$ is minimally coupled to the metric, but its kinetic term is the standard kinetic term $X$ raised to a power involving the model parameter $\cq$. The constant $\Lambda$ is not the cosmological constant here, it is another model parameter with the same dimensions as $X$, and its value is important for ensuring that $Q$ behaves as dark energy at late times. This dark energy model was considered in \cite{Garriga:1999vw, Gordon:2004ez}, and was shown to contain quintessence perturbations that are conserved outside the horizon. These dark energy perturbations act as a source of growing isocurvature modes in cold dark matter. 

	In the uniform field gauge (which is also the comoving gauge), where perturbations in $Q$ vanish, the sound speed is the constant $c_Q$. For $c_Q=1$, the kinetic term in Eq.~(\ref{eq:L}) reduces to the canonical one. If this sound speed is chosen to be small compared to the speed of light, the dark energy perturbations cluster after horizon crossing. 

	The equation of state parameter, $\wq$, can be fixed by choosing an appropriate potential. However, if $\wq$ is chosen to be constant, then the adiabatic sound speed is
	\begin{align}
		\ca \equiv \frac{\dot{\bar{p}}_{{}_Q}}{\dot{\bar{\rho}}_{{}_Q}} = \wq \, ,
	\end{align}	
	where $\bar{\rho}_Q$ and $\bar{p}_Q$ are the energy density and pressure of the homogenous, background quintessence field. The dots denote derivatives with respect to time. As we will see in the following, gravitational coupling between the matter and quintessence perturbations makes it  impossible to have conserved isocurvature on superhorizon scales under these conditions. The back reaction of matter onto quintessence will cause it to evolve even on superhorizon scales. 

	To avoid this, the equation of state can instead be fixed by choosing $\Lambda$ and $V(Q)$ so that the potential is large in comparison with the kinetic term. In this way, the equation of state is held at $\wq\simeq-1$ by the large value of the potential, while the adiabatic sound speed depends only on the slope of the potential.

	The equation of motion for the homogeneous background field is
	\begin{align}
		Q'' + \lb 3\cq + \frac{H'}{H} \rb Q' + \frac{\cq V\!\!,_{Q}}{H^2} \lb \frac{H^2 Q'{}^2}{2\Lambda} \rb ^ { \frac{\cq -1}{2\cq} } = 0 \, ,
	\end{align} 	
	where $H = H(a)$ is the Hubble rate and primes are derivatives with respect to $\log(a)$. 

	For simplicity we will assume the potential has a constant slope $V\!\!,_{Q}$, which allows us to find an exact solution,
	\begin{align}
		\label{eq:Qp}
		Q' & = -\sqrt{\frac{2\Lambda}{H^2}} \lb \frac{V\!\!,_{Q}} { \sqrt{2\Lambda H^2}} f(a) \rb^{\cq} \, , \\
		\label{eq:f}
		f(a) & =\frac{H(a)}{a^3} \left( \int^a_{a_i} \frac{dx}{x} \frac{x^3}{ H(x)} + \frac{a_i^3}{H(a_i)} f(a_i) \right)\, .
	\end{align}
	The integration constant $f(a_i)$ fixes the value of $Q'$ at some initial time $a_i$. Due to the factor of $a^{-3}H(a)$ in Eq.~(\ref{eq:f}), and the small value of the sound speed, the integration constant is unimportant relative to the first term in $f(a)$ so we set $f(a_i)$ to zero. In particular, during inflation the solution for $Q'$ rapidly evolves to a constant that is independent of the initial condition. The slope of the potential is not required to be small, since the change in the field strength is determined by the constant $\Lambda$.

	\begin{figure*}
		\includegraphics[width=\linewidth]{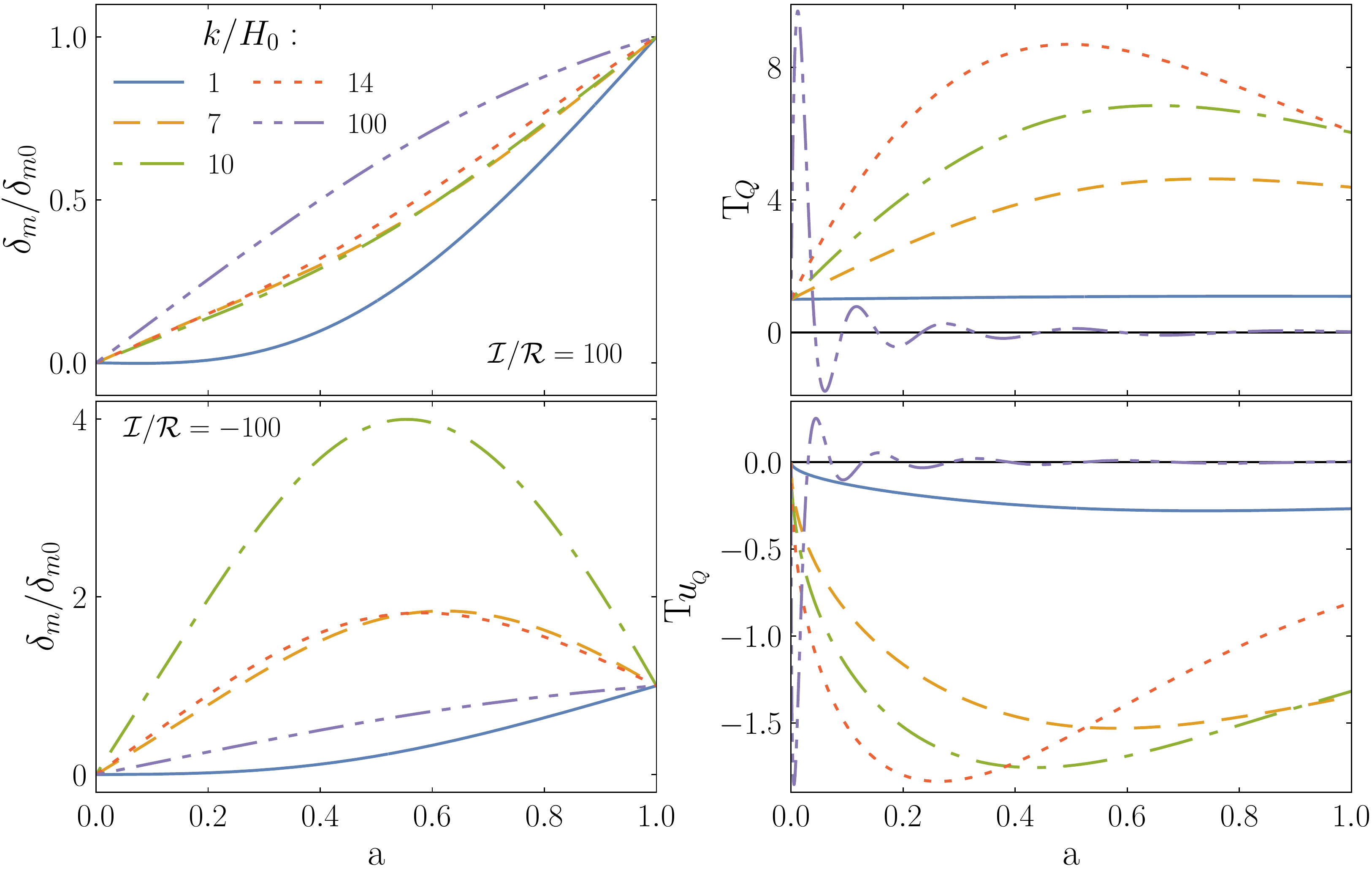}
		\caption{Left: evolution of long-wavelength matter perturbations, normalized to final value 1, sourced by both curvature and quintessence isocurvature for positively correlated (top) or anticorrelated (bottom) $\I/\R$.  Right: transfer functions for the quintessence energy density contrast (top) and velocity perturbation (bottom). The magnitude of the ratio of primordial isocurvature to curvature is fixed to 100 to emphasize the effect of the isocurvature mode. The sound speed choice $c_Q=0.1$ puts the Jeans scale at $k_{J}/H_0=10$ today. The evolution of the $\delta_m$ mode with $k/H_0=100$ is indistinguishable from the evolution of a purely adiabatic mode of any wavelength. The legend in the top left applies to all plots.}
		\label{fig:d}
	\end{figure*}

	Integrating Eq.~(\ref{eq:Qp}) gives the background evolution of the quintessence field. However, by assumption, the background field is dominated by a large integration constant $Q_0$. For a given potential, $V(Q_0)$ is fixed to reproduce the present day energy density of the universe. The constant $\Lambda$ is then used to ensure that the time dependent term in $Q$ is always small compared to $Q_0$. This is different from the typical slow-roll assumptions, which enforce both that $w\simeq-1$, and that the fractional change in $w$ over a Hubble time is small. In our case, the fractional change in the equation of state is not small during eras when $H'/H$ is not small, but the equation of state was so close to $w=-1$ initially that $Q$ still behaves as dark energy today.

	The adiabatic speed of sound for this solution is
	\begin{align}
		\ca = \cq - \frac{1+\cq}{3} \frac{1}{f(a)} \, .
	\end{align}
	At early times during matter domination, for $\cq \ll 1$, this gives $\ca \simeq -3/2$. Note, the adiabatic sound speed is only the sound speed of physical perturbations if they are adiabatic. For stability, the negative value of $\ca$ indicates that there must nonadiabatic stress, and therefore entropy perturbations.

	The equations for linear perturbations are convenient to analyze in the synchronous gauge \cite{PhysRevD.22.1882,Kodama:1984PTPS}. Linear perturbations in the scalar components of the quintessence stress-energy tensor are gauge-dependent quantities. In the uniform $Q$ gauge and synchronous gauge, they are related by
	\begin{align}
		\delta\rho_{Q}\big|_{\rm{u}} & = \delta\rho_Q - \bar{\rho}_Q\,\!\!\!\!'\ \frac{aH}{k}\frac{\uq}{1+\wq} \, , \\
		\delta p_{Q}\big|_{\rm{u}}  & = \delta p_Q -  \bar{p}_Q\,\!\!\!\!'\ \frac{aH}{k}\frac{\uq}{1+\wq} \, .
	\end{align}
	Here, $\delta\rho_{Q}$ is the linear perturbation to the quintessence energy density, $\delta p_Q$ is the  linear perturbation to the its pressure, and $u_Q$ is the velocity potential of the quintessence field, all evaluated in the synchronous gauge. The quantities on the left are evaluated in the uniform field gauge, which is  comoving with $Q$, so the velocity potential $u_Q|_{\rm{u}}$ vanishes. 

	From the above transformation, we can obtain the sound speed in the synchronous gauge, which is
	\begin{align}
		\label{eq:cs}
		\cs \delta_Q \equiv \frac{\delta p_Q}{\bar{\rho}_Q} = \cq\dq + 3 \lb \cq - \ca \rb \frac{a H}{k}\uq \, .
	\end{align}
	The equation of state and synchronous gauge sound speed can be used to eliminate the pressure and its linear perturbation from the quintessence continuity and Euler equations. The continuity equation for the pressureless matter can be used to eliminate the scalar metric degree of freedom. The resulting system of differential equations describing the linear perturbations of the combined matter-quintessence cosmic fluid is
	\begin{align}
		\dq' + 3\lb \cs - \wq \rb \dq =  & - \frac{k}{a H} \uq + \lb 1 + \wq \rb \dm' \, , \\
		\uq' + (1 -3\wq)\uq = & \frac{k}{a H} \cs \dq\, , \\
		\dm'' + \lb 2 +  \frac{H'}{H} \rb \dm' = & \frac{3}{2}\frac{H_0^2}{H^2}\frac{\om}{a^3}\dm \\
		 & + \frac{3}{2}\frac{H_0^2}{H^2}\frac{\oq}{a^{3(1+\wq)}} \lb 1 + 3\cs \rb \dq \, . \nonumber
	\end{align}
	Energy density contrasts $\delta_i$ are defined
	\begin{align}
		\delta_i = \frac{\delta\rho_i}{\bar{\rho}_i} - 1 \, ,
	\end{align}
	with $i=m$ for matter or  $Q$ for quintessence. 

	If $\wq\neq-1$, then there are no scaling solutions, $\delta_i \propto a^{\gamma_i}$ for some constant exponents $\gamma_i$. In this case, the quintessence perturbations evolve outside the horizon. Alternatively, if $\wq=-1$, then the matter does not appear in the quintessence continuity equation, and the condition for constant superhorizon quintessence becomes $\cs=-1$ for $k\ll aH$. In the limit where $\cq\ll1$, the initial conditions for these perturbations during matter domination are
	\begin{align}
		\label{eq:dq}
		\dq(a_i,k) & = \I(k) \, , \\
		\label{eq:icu}
		\uq(a_i,k) & = -f(a_i)\frac{k}{a_i H_i} \I(k) \, , \\
		\label{eq:dm}
		\dm(a_i,k) & = \frac{2}{5}\lb\frac{k}{a_i H_i}\rb^2\R(k) -\frac{1}{3}\frac{\oq}{\om}a_i^3\I(k) \, .
	\end{align}	
	where $f(a_i) = 2/9$ during matter domination. The matter perturbations are sourced by both primordial isocurvature fluctuations ($\I$) and curvature fluctuations ($\R$). The latter correspond to the homogeneous solution of the linear growth equation for matter.

	Introducing transfer functions for the quintessence energy density contrast, velocity potential, and the separate curvature and isocurvature components of the matter density contrast
	\begin{align}
		\dq(a,k) & = T_{Q}(a,k)\I(k) \, , \\
		\uq(a,k) & = T_{u_{Q}}(a,k)\I(k) \, , \\
		\dm(a,k) & = \tmr(a,k)\R(k) + \tmi(a,k)\I(k) \, .	
	\end{align}
	The total linear matter power spectrum is then
	\begin{align}
		P_{mm}(k) = &\ (\tmr(a,k))^2P_{\R\R}(k) \\
		& + 2 \tmr(a,k)\tmi(a,k) P_{\R\I}(k)\nonumber \\ 
		& + (\tmi(a,k))^2P_{\I\I}(k) \nonumber\\
		\equiv &\ P_{mm}^{\R\R}(k)+2 P_{mm}^{\R\I}(k) + P_{mm}^{\I\I}(k) \, ,
	\end{align}
	where $P_{\R\R}$, $P_{\R\I}$ and $P_{\I\I}$ are the auto and cross-power spectra for primordial curvature and isocurvature. In the last line we have defined $P_{mm}^{XY}$ as parts of the matter perturbation sourced by curvature and isocurvature spectra. On sub-Jeans scales, the total matter power spectrum is just the contribution from adiabatic perturbations, $P_{mm}^{\R\R}$. In this regime we will always choose $P_{mm}^{\R\R}(k)$ to be the usual $\Lambda$CDM matter power spectrum (our specific parameter choices will be given in Table~\ref{tab:cos}. On larger scales, we leave the $P_{\R\R}$, $P_{\R\I}$ and $P_{\I\I}$ unspecified, but in Section~\ref{sec:observables} we will show how different assumptions about the power spectra impact observables.

	The curvature-sourced transfer function depends on $k$ only through its initial condition. The isocurvature-sourced transfer function has $k$-independent initial conditions, but it has $k$-dependent evolution. This scale dependence ultimately originates in the velocity gradient in the quintessence continuity equation, which becomes important after horizon crossing. Scale-dependent growth occurs from the horizon down to the Jeans scale, below which the quintessence perturbations are pressure supported so they do not grow. Far below the Jeans scale, the matter is dominated by its adiabatic component with scale-independent growth.

	Solutions to the linear growth equations are plotted in FIG. \ref{fig:d}, with initial conditions fixed by setting the final $\dm{}_{0}$  to be the same value for all matter modes. To emphasize the effects of isocurvature we fix the ratio of primordial curvature to isocurvature to be $\I/\R=\pm100$ outside the horizon. The comoving sound speed is set to $c_Q = 0.1$, which fixes the quintessence Jeans scale today to be $k_J \simeq 10\ H_0$. Both the curvature and quintessence fluctuations are constant outside the horizon. After horizon crossing, the quintessence grows until crossing the Jeans scale. Below the Jeans scale, these perturbations oscillate and quickly decay away, so $k\simeq100\ H_0$ is essentially a purely adiabatic mode.

	During matter domination, the matter components sourced by curvature and isocurvature grow as $a$ and $a^3$ respectively. There are two scale-invariant regimes for matter growth, the isocurvature-dominated regime at large scales and the curvature-dominated regime at small scales. The ratio $\I/\R$ and the quintessence Jeans scale determine where the transition is between these two regimes. The larger $\I/\R$ becomes, the closer the transition is to the quintessence Jeans scale. The scale dependent growth is most extreme near these intermediate scales.

	For the matter perturbations, the behavior depends on the relative sign of the primordial curvature and isocurvature fluctuations. Since the two terms in the initial condition for the matter have opposite signs, correlated curvature and isocurvature tend to decrease the matter perturbations below their adiabatic value, whereas they are increased by anticorrelated primordial fluctuations. The net effect is that matter perturbations grow rapidly at late times in the correlated case, while they approach a maximum value and then decrease when $\R$ and $\I$ are anticorrelated. 
	
\section{Separate universe}
\label{sec:SU}
		
	The following summarizes previous work by Hu \emph{et al.} \cite{Hu:2016ssz}, in which separate universe techniques were developed for large-scale structure in the presence of a long-wavelength Jeans scale. In the separate universe formalism, we consider a region embedded in a long-wavelength matter perturbation $\dl$, which is approximately spatially homogeneous across the region. In this region, the long-wavelength perturbation appears as a shift in the average matter energy density,
	\begin{align}
		\bar{\rho}_{mW}(a;k) = \bar{\rho}_{m}(a)\lb 1 + \dl(k,a) \rb \, .
	\end{align}
	Here, the local quantities in the ``windowed'' region are denoted with a subscript $W$.  The wave number of a long-wavelength perturbation under consideration is $k$.  We restrict our analysis to windowed regions that are smaller than the quintessence Jeans scale so that the quintessence perturbations can be ignored, and the only fluctuations around $\bar{\rho}_W$ are fluctuations in matter. Wave numbers of modes smaller than the size of this region, and therefore smaller than the quintessence Jeans scale, will be denoted below as $k_S$.
	
	Requiring that the local matter density evolves as $a_W^{-3}$ with respect to the local scale factor gives
	\begin{align}
		\frac{\om H_0^2}{a^3}(1+\dl) = \frac{\om{}_W H_{0W}^2}{a_W^3} \, .
	\end{align}
	Since the long-wavelength perturbation is negligible at early times, the two cosmologies initially coincide. At leading order in $\dl$, this condition fixes the transformation from the global to the local cosmology,
	\begin{align}
		\om{}_W H_{0W}^2 & = \om H_0^2 \, , \\
		\label{eq:t1}
		a_W & \simeq a \lb 1- \frac{1}{3} \dl \rb  \, , \\
		\label{eq:t2}
		H_W & \simeq H \lb 1 - \frac{1}{3} \dl' \rb  \, , \\
		\label{eq:t3}
		\frac{d}{d\log \! a_W} & \simeq \lb 1 + \frac{1}{3}\dl' \rb \frac{d}{d\log\! a} \, .
	\end{align}
	
	\subsection{Linear perturbations}
	
	Consider a small-scale mode of the local matter perturbation $\delta_{mW}$. The evolution of this mode with respect to the local cosmology satisfies the same growth equation as adiabatic matter perturbations with respect to the global cosmology. For linear perturbations, we write $\delta_{mW}$ in terms of a local linear growth factor,
	\begin{align}
		\dmw(\ks,a;k) = D_{W}(a;k) \dm(\ks,a_i) \, ,
	\end{align}
	where $k$ is the wave number of long-wavelength mode, and we are choosing $D_W(a_i;k)=1$. If the initial condition for $\dmw$ at scale factor $a_i$ is set early enough, its dependence on the long-wavelength mode is negligible. The evolution of the linear growth factor is given by
	\begin{align}
		\frac{d^2 D_{W}}{d\log\! a_W{}^2} + \lb 2 + \frac{d\log\! H_W}{d\log\! a_W} \rb & \frac{d D_{W}}{d\log\! a_W} \nonumber \\
		\label{eq:bgdm}
		 -  \frac{3}{2}&\frac{H_{0W}^2}{H_W^2}\frac{\om{}_W}{a_W^3} D_{W} = 0 \, .
	\end{align} 
	Small-scale perturbations are well below the Jeans scale, so at these scales the quintessence perturbations are negligible, which is why they do not show up on the right-hand side of the above expression. The local matter growth factor can be decomposed into a term equal to the global growth factor and a response term sourced by the long-wavelength mode,
	\begin{align}
		\label{eq:dme}
		D_{W}(k,a) = D(a) \Big( 1 + \eps(a;k) \Big) \, .
	\end{align}	
	Using the transformations in Eqs.~(\ref{eq:t1})--(\ref{eq:t3}) to rewrite Eq.~(\ref{eq:bgdm}) in terms of the global cosmology gives
	\begin{align}
			D'' + \lb 2 + \frac{H'}{H} \rb D' & = \frac{3}{2}\frac{H_{0}^2}{H^2}\frac{\om{}}{a^3}D \, , \\
			\eps'' + \lb 2 + 2\frac{D'}{D} + \frac{H'}{H} \rb\eps' & =  \frac{3}{2}\frac{H_{0}^2}{H^2}\frac{\om{}}{a^3}\dl + \frac{2}{3}\frac{D'}{D}\dl' \, .
	\end{align}
	The small-scale mode is only sourced by curvature, but the long-wavelength mode has both curvature and isocurvature contributions. We can define transfer functions for the response, 
	\begin{align}
		\label{eq:te}
		\eps(a;k)= \ter(a;k)\R(k) + \tei(a;k)\I(k) \, .
	\end{align}
	 During matter domination, the initial conditions are
	\begin{align}
		\ter(a_i;k) & = \frac{13}{21}\tmr(a_i;k)\, , \\
		\tei(a_i;k) & =  \frac{7}{33}\tmi(a_i;k)\, .
	\end{align}
	
	\subsection{Power spectrum growth response}
	
	\begin{figure*}
		\includegraphics[width=\linewidth]{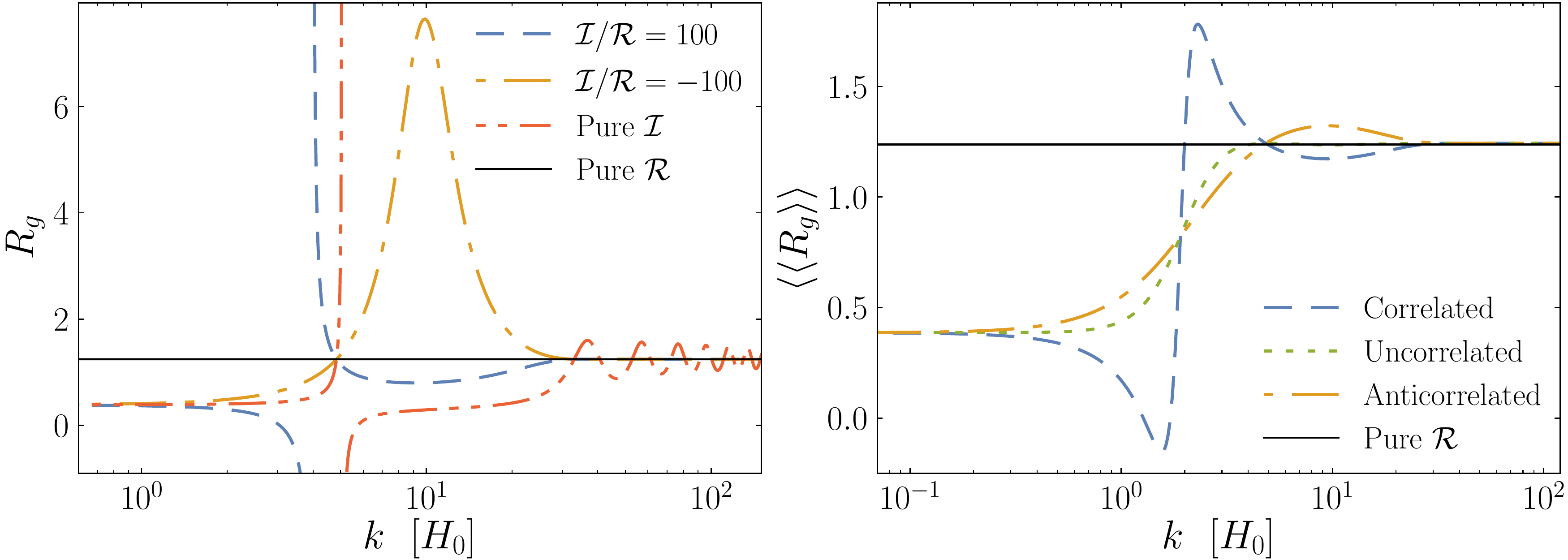}	
		\caption{Left: the $z=0$ power spectrum response as a function of large-scale wave number for a single realization of the long-wavelength mode, with $\I/\R=\pm100$. The divergence in the correlated case ($\I/\R>0$) is due to the initial condition for the curvature and isocurvature components of the matter perturbations having opposite sign. There is a scale at which the two components cancel causing the total matter perturbation to vanish.  Right: the $z=0$ power spectrum response assuming scale invariant power spectra for the primordial curvature and quintessence isocurvature perturbations. The cross-correlation is taken to be either vanishing, or $\pm90\%$ the Cauchy-Schwarz bound, with isocurvature amplitude 100 times larger than the curvature amplitude. The double bracket indicates averaging in the following sense: $\braket{\braket{R_g}} = \braket{R_g\dm\dm}/\braket{\dm\dm}$. }	
		\label{fig:Rs}
	\end{figure*}
	
	The local matter-matter power spectrum of a region within a long-wavelength mode will differ from the global power spectrum. The difference can be characterized by three contributions: change in comoving wavelengths, change in the definition of the average background energy density, and change in the local growth factor \cite{Takada:2013bfn}. These contributions are summarized by
	\begin{align}
		\label{Eq:dlnPddl}
		\frac{d\log P_{mmW}}{d\dl} = R_d + R_{\bar{\rho}} + R_g \, .
	\end{align}
	The first term refers to the dilation of comoving wavelengths, which is due to the local scale factor's dependence on the long-wavelength mode. This term can be calculated
	\begin{align}
		R_d = -\frac{1}{3} \frac{d\log\! \lb P_{mm} \rb}{d\log \ks} \, .
	\end{align}
	The second term refers to the shift in the mean energy density, which changes the definition of energy density contrasts. This is simply
	\begin{align}
		R_{\bar{\rho}} = 2 \, .
	\end{align}
	The final term, referring to change in the growth factor due to the long-wavelength mode, can be estimated by taking the finite difference derivative between the local power in overdense and underdense separate universe regions,
	\begin{align}
		\label{Eq:Rg}
		R_g = \frac{P_{mmW}(a,\ks|+\dl)- P_{mmW}(a,\ks|-\dl)}{2P_{mm}(a,\ks)\dl} \, .
	\end{align} 
	Note, this last term is the only contribution to the power spectrum response that depends dynamically on the long-wavelength mode. The other two contributions can be calculated without separate universe simulations. All of the references to the power spectrum response below refer only to the growth part of the response. 
	
	Since $R_g$ depends dynamically on the long-wavelength mode, it will depend on the evolutionary history of $\delta_L$ \cite{Ma:2006zk}. Therefore $R_g$ depends on the particular ratio of adiabatic and isocurvature modes that comprise $\delta_L$. For a single realization, $\delta_L = \tmr \R + \tmi\I$, the power spectrum response is given by
	\begin{align}
		\label{Eq:RgofTransfers}
		R_g & = 2 \frac{\ter \R + \tei\I}{\tmr \R + \tmi \I} \, .
	\end{align}
   From Eq.~(\ref{Eq:RgofTransfers}) it is clear that the total growth response in Eq.~(\ref{Eq:Rg}) from a single realization of $\dl$ is in general a random variable, dependent on the realization of $\I$ and $\R$. This is in contrast to a cosmology in which a single degree of freedom determines $\dl$ and each mode of $\delta_L(k)$ has the same evolutionary history (e.g., for adiabatic perturbations in $\Lambda$CDM \cite{Wagner:2014aka, Dai:2015jaa, Hu:2016ssz}, quintessence \cite{Chiang:2016vxa}, or neutrino-CDM cosmologies \cite{Chiang:2017vuk}).

    In what follows it will be helpful to define individual growth responses for the pieces of $\dl$ generated by $\I$ and $\R$,
	\begin{align}
		R_g^\R &= 2\frac{\ter}{\tmr} \, , \\
		R_g^\I &= 2\frac{\tei}{\tmi} \, .
	\end{align}
	Unlike the total growth response in Eq.~(\ref{Eq:RgofTransfers}), the individual responses to the adiabatic and isocurvature terms are not realization-dependent. The existence of two modes, $\I$ and $\R$, changes the usual separate universe relationships between $\epsilon$, $R_g$, and the squeezed-limit bispectrum. Determining the change in the power spectrum due to the presence of a long-wavelength mode from Eq.~(\ref{Eq:dlnPddl}) requires knowledge of the local values of $\R$ and $\I$ that generated the long-wavelength mode. As we shall see, the squeezed-limit bispectrum will depend on $P_{\R\R}$, $P_{\R\I}$, and $P_{\I\I}$. 	
	
	The power spectrum response for single realizations of $\dl$ is plotted in FIG. \ref{fig:Rs} as a function of the large-scale wave number. Far above the Jeans scale, the response is isocurvature dominated and scale invariant. Far below the Jeans scale, the response approaches the purely adiabatic value. Between the horizon crossing scale and Jeans scale the response is scale dependent. In the correlated case ($\I/\R>0$), the initial conditions for the curvature and isocurvature components of the matter modes have opposite sign. At a given redshift there is a scale at which the two components cancel, so the total matter perturbation vanishes, causing the response to diverge at this scale. 
	
	The squeezed-limit bispectrum can be determined by computing the correlation between the local power spectrum and $\dl$, $\langle P_{mmW}(a,\ks |\dl) \dl\rangle$. Computing the ratio we find
    \begin{align}
    	\frac{\langle P_{mmW}(a,\ks |\dl) \dl\rangle}{P_{mm}(k_S)\braket{\delta_L\delta_L}} = & \braket{ \frac{d \log P_{mmW}}{d\dl}\frac{\dl\dl}{\braket{\delta_L\delta_L}}} \, , \\
    	 \label{eq:Bsq}
    	= &\ R_d + R_{\bar{\rho}}  \\
    	& + R_g^\R \left( \frac{P_{mm}^{\R\R}(k) + P_{mm}^{\R\I}(k)}{P_{mm}(k)} \right) \nonumber\\
   		& + R_g^\I  \left(\frac{P_{mm}^{\I\I}(k) + P_{mm}^{\R\I}(k)}{P_{mm}(k)} \right)\nonumber\, .
    \end{align}
	The coefficients of the growth responses depend on the primordial power spectra, $P_{\R\R}$, $P_{\R\I}$, and $P_{\I\I}$, which depend on the mechanism, inflationary or otherwise, that generates these fluctuations in the early universe. Here we make no attempt to find a model that would generate these primordial fluctuations in the early universe. However, the \hyperref[sec:generationmodes]{Appendix} at the end of this paper considers the viability of certain models that could produce isocurvature modes that are significantly correlated or anticorrelated with the primordial curvature modes. 
	
	The last two terms in Eq.~(\ref{eq:Bsq}) can be interpreted as the growth part of the power spectrum response averaged over the long-wavelength mode in the following sense:
	\begin{align}
		\braket{\braket{R_g}} = \frac{\braket{R_g\delta_L\delta_L}}{\braket{\delta_L\delta_L}} \, .
 	\end{align}
	For the purposes of illustrating the effects of averaging over long-wavelength modes, we make the assumption that both of the auto power spectra and the cross power spectrum of the primordial curvature and isocurvature fluctuations are scale invariant. FIG. \ref{fig:Rs} shows examples of the averaged power spectrum responses with the isocurvature amplitude taken to be 100 times larger than the curvature amplitude to emphasize its effect. Examples are shown with cross-correlation vanishing, and with the cross-correlation $\pm90\%$ its bound with respect to the Cauchy-Schwartz inequality, $|P_{\R\I}|\leq\sqrt{P_{\R\R}P_{\I\I}}$.
	
	Responses of small-scale observables to a long-wavelength mode that are described by the separate universe formalism are closely related to angle-averaged, equal-time cosmic consistency relations. The presence of quintessence isocurvature violates the assumptions used to derived the standard cosmic consistency relations \cite{Valageas:2013zda}. The expression for the squeezed-limit bispectrum shown in Eq.~(\ref{eq:Bsq}) is a generalization of the bispectrum angle-averaged, equal-time cosmic consistency relation to the case of quintessence isocurvature. 
	
\section{Simulations}
\label{sec:sims}

	Separate universe simulations involve computing the evolution of an N-body system of particles within the context of a long-wavelength mode that is treated as a shift in the homogeneous background energy density.  In our case, the particles are cold dark matter and baryons, which are treated equivalently because we are interested in scales much larger than the baryonic Jeans scale. The system includes $N_p=(512)^3$ nonrelativistic particles interacting only through Newtonian gravity on an Friedmann-Robertson-Walker expanding background. The
	comoving size of the box is fixed locally to length $L_W = 500\ \rm{Mpc}$$/h $. 
	
	The expansion is characterized by the local Hubble rate $H_W$ as a function of local scale factor $a_W$, which differs from $\Lambda$CDM by a perturbatively small contribution due to the presence of the long-wavelength mode $\dl$. At a given wave number, $\dl$  is computed by numerically integrating Eqs.~(\ref{eq:dq})--(\ref{eq:dm}). Values of the local scale factor and Hubble rate are tabulated using Eqs.~(\ref{eq:t1})--(\ref{eq:t3}). The background expansion is fixed entirely by $\dl$ and the choice of global cosmology parameters $\om=0.3$, $\oq=0.7$, and $\wq=-1$.
	
	The N-body simulations were run using the code Gadget2 \cite{Springel:2005mi}, modified to read in tabulated values of the separate universe scale factor and Hubble rate for a given long-wavelength mode. Sets of 20 simulations were run for the long-wavelength modes $k/H_0$=1,  7, 10, 14, and 100 with $\I/\R=-100$, and $k/H_0$=1, 10, and 100 with $\I/\R=100$. These range from the isocurvature to the curvature dominated regime, capturing features of the scale dependent power spectrum response. Additional simulations were run for a pure isocurvature mode at $k/H_0=10$, and a pure curvature mode. With both overdense and underdense separate universe simulations for each mode, 400 separate universe simulation were run in total. An additional 20 were from for a  larger global universe box of size $L=1000\ \rm{Mpc}/h$ with $(512)^3$ particles with standard adiabatic initial conditions. These were used to measure the clustering bias as a check on our separate universe results for the adiabatic response bias.
	
	Initial conditions were generated as realizations of Gaussian random fields from the matter power spectrum at the initial simulation time $z_i=49$. Although the baryons behave indistinguishably from the dark matter in the N-body interactions, their presence affects the shape of the power spectrum through the baryonic acoustic oscillations. To account for this, the power spectrum is first calculated at redshift $z=0$ with the code CLASS \cite{Blas:2011rf}, using the cosmological parameters given in Table~\ref{tab:cos}. The matter power spectrum is then rescaled back to the initial simulation time. At early times, the difference between local and global cosmology is negligible, so the primordial fluctuations are the same for both. However, the linear growth factors are not the same. To account for this, we rescale the power spectrum back using the global growth factor and forward to the initial simulation time using local scale factor
	\begin{align}
		P_{W}(\ks, a_{Wi};k) = P(\ks, a_0)\lb \frac{D_W(k,a_{Wi})}{D(a_0)} \rb^2 \, .
	\end{align}
	For each long-wavelength mode, the same 20 random seeds were used to generate the initial conditions for the particles' positions and velocities. These were corrected using a second order Lagrangian perturbation theory code to reduce transients \cite{Crocce:2006ve}. Simulation snapshots were taken at values of the global redshifts $z=1.0,\ 0.75,\ 0.5,\ 0.25,\ \mathrm{and}\ 0.0$. Since the simulations are run in local time, their snapshot output times are adjusted according to Eq.~(\ref{eq:t1}) to make sure overdense and underdense boxes are matched at the same the global time.
	
	\begin{table}
		\begin{tabular}{c c}
			\hline\hline
			Parameter & Value \\
			\hline
			$\oq$ & 0.7 \\			
			$\om$ & 0.3 \\
			$\ob$ & 0.05 \\
			$h$ & 0.7 \\
			$n_s$ & 0.968 \\
			$A_s$ & 2.137$\times10^{-9}$ \\ 
			$N_p$ & $(512)^3$ \\
			$L_W$ & $500\ \rm{Mpc}/$$h$ \\
			$M_p$ & $1.108\times10^{11}\ \rm{M}_{\odot}$\\
			\hline
			\hline
		\end{tabular}
		\caption{Cosmological and N-body simulation parameters.}
		\label{tab:cos}
	\end{table}

	\begin{figure}
		\includegraphics[width=\linewidth]{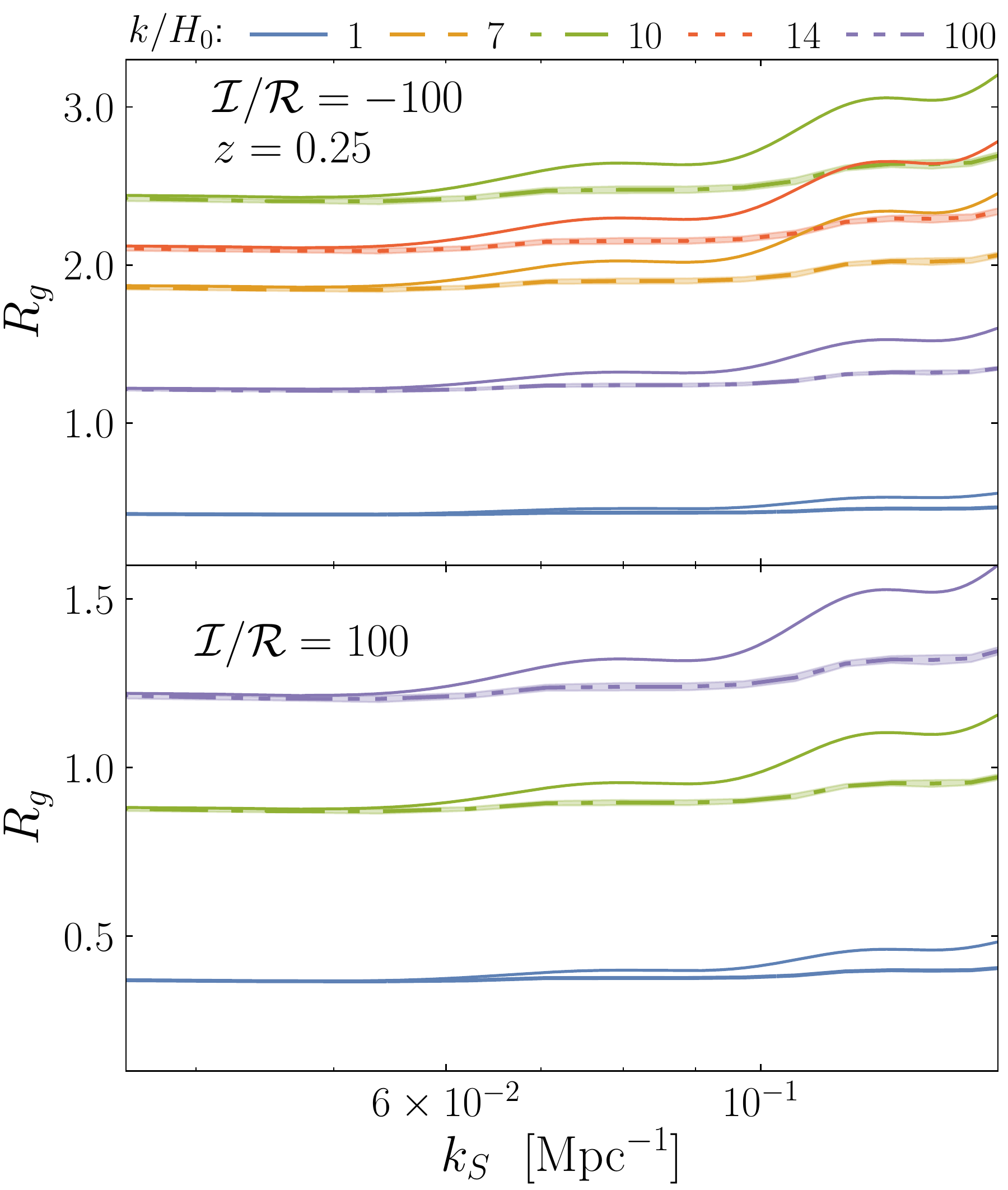}	
		\caption{Power spectrum response at $z=0.25$ from separate universe simulations. The shaded in regions show the $1$-$\sigma$ bootstrap error, the thin solid lines show the 1-loop calculation from standard perturbation theory. We note that the ratios of the different $R_g$ to the adiabatic growth response are constant with respect to $k_S$.}
		\label{fig:dP}
	\end{figure}
	
\section{Power spectrum response}
\label{sec:Presponse}
		
	For a given simulation snapshot, the particle positions were converted to a density field, which was Fourier transformed using FFTW
	\cite{FFTW05}. The density field was estimated by distributing the particles to sites on a $(1024)^3$ grid, using the cloud-in-cell method. The power spectrum was then calculated from the Fourier transformed density field, and the power spectrum response was estimated through the finite difference derivative in Eq.~(\ref{Eq:Rg}), from the overdense and underdense separate universe simulations. By resampling over the 20 random realizations, the bootstrap variance in the power spectrum response was calculated.
	
	Effects of weakly nonlinear growth can be captured by calculating the power spectrum using perturbation theory. At 1-loop we have \cite{Jeong:2006xd}
	\begin{align}
		P_{1\dsh\rm{loop}} = P_{11} + P_{22} + 2P_{13} \, ,
	\end{align}
	where the first term is the linear power spectrum and last two terms are proportional to $\lb D_W\rb^4$. The 1-loop correction to the power spectrum response is
	\begin{align}
		R_{g,1\dsh\rm{loop}} = \lb 1 + \frac{P_{22}+2P_{13}}{P_{11}} \rb R_g \, .
	\end{align}
	Here $R_g$ is the linear power spectrum response. The power spectrum responses measured from our simulations at $z=0.25$ are plotted in FIG. \ref{fig:dP},  along with a 1-loop calculation shown for each value of $k$ that was simulated. Above the nonlinear scale, the response measured from the simulations agrees well with the linear power spectrum response. Agreement with the 1-loop calculation continues until $k_S\simeq6\times10^{-2}\ \rm{Mpc}^{-1}$, which is approaching the nonlinear scale where perturbation theory is no longer valid.
	
\section{Halo bias}
\label{sec:bias}

	\begin{figure*}
		\includegraphics[width=\linewidth]{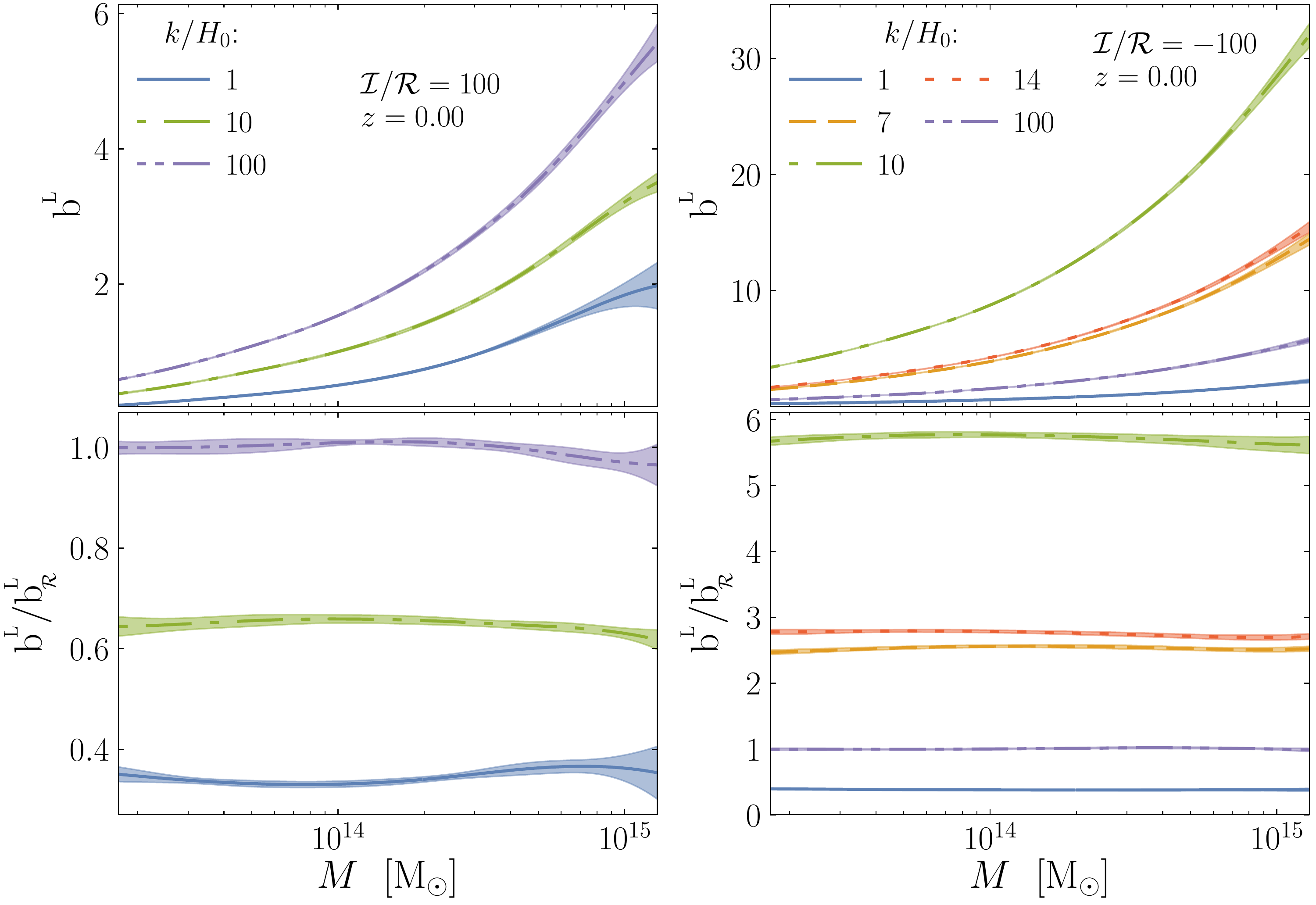}
		\caption{Top: Lagrangian bias for each long-wavelength mode at redshift $z=0$.  Bottom: ratio of Lagrangian bias to the purely adiabatic Lagrangian bias. Correlated and anticorrelated cases are on the left and right respectively. The shaded regions show the variance estimated from bootstrap resampling.}	
		\label{fig:bs}
	\end{figure*}

	\begin{figure*}
		\includegraphics[width=0.49\linewidth]{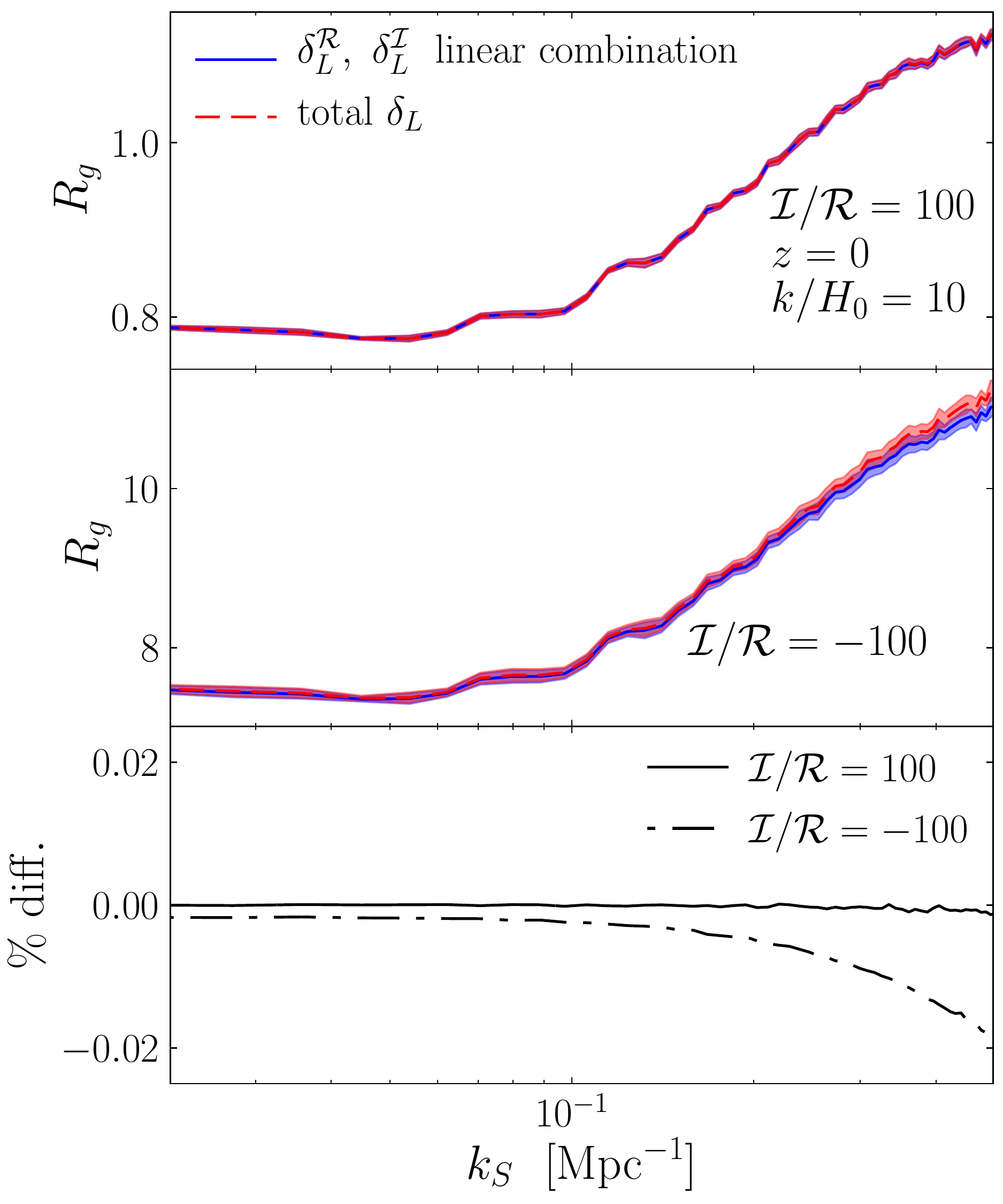}
		\includegraphics[width=0.49\linewidth]{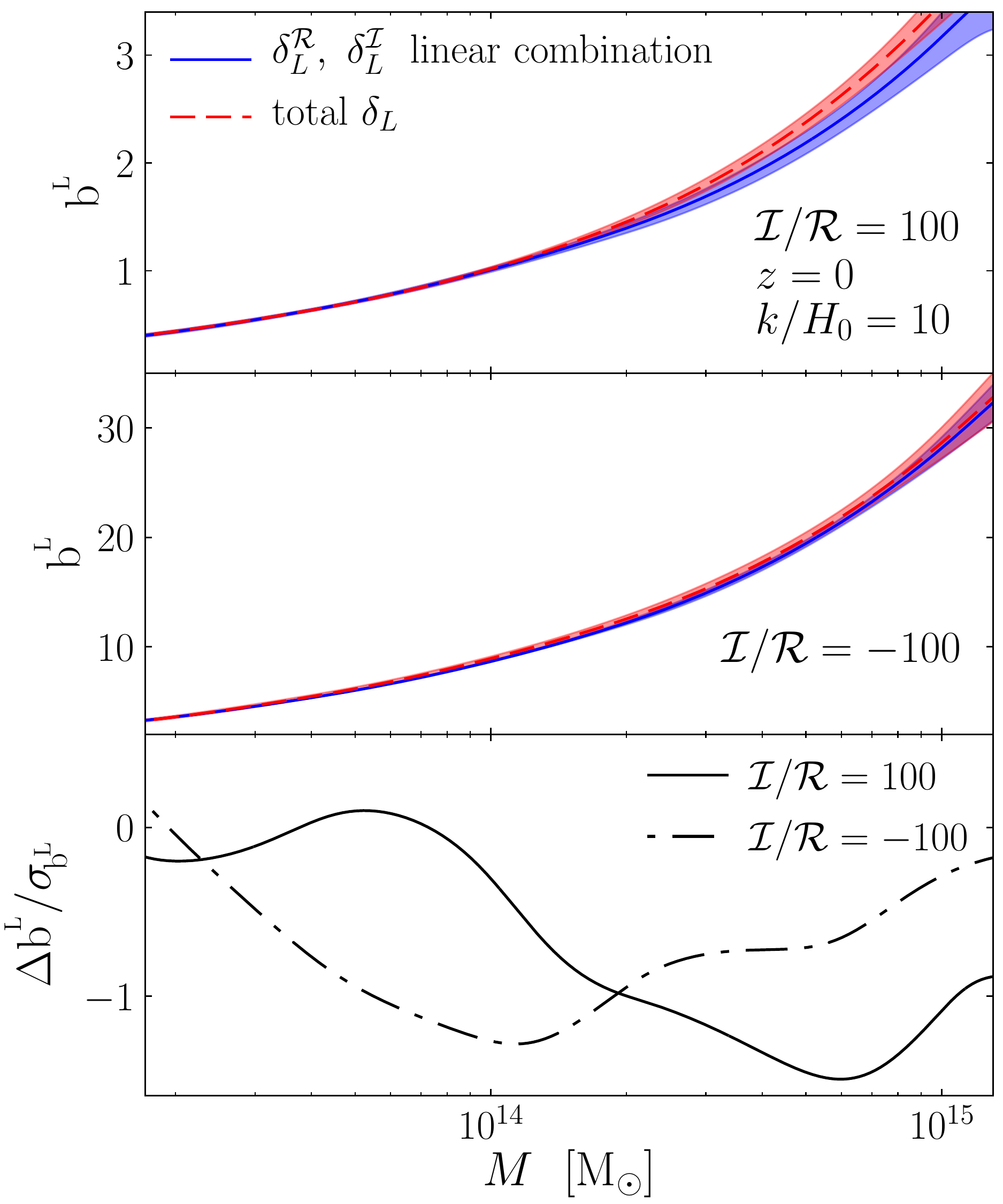}
		\caption{Linear sum of simulation results for individual components of the long-wavelength matter perturbation compared with the total matter perturbation results. The left plot shows the power spectrum responses, and the bottom shows the percentage difference, which is consistent with numerical errors from taking finite difference derivatives. The right plot shows the Lagrangian bias. The bottom shows the number of standard deviations between the linear combination and the total matter perturbation determinations of the bias.}	
		\label{fig:lin}
	\end{figure*}

	The Lagrangian bias was determined for each simulation set by using the abundance matching method \cite{Li:2015jsz}, which involves calculating the mass shift at a fixed cumulative number density between overdense and underdense separate universes. For a given set of simulations, a cumulative catalog of halo masses, from all 20 realization of the initial conditions, was constructed and then sorted in descending order. These sorted mass catalogs give an estimate of the cumulative number density of objects above some threshold mass,
	\begin{align}
		n(\log\! M_{th};\dl) = \int_{\log\! M_{th}}^\infty d\log\! M \lb-\frac{d n^L(\log\! M, \dl)}{d\log\!M} \rb\, .
	\end{align}
	From our sorted halo catalogs, we construct the following lists:
	\begin{align}
		\log\! M_i & = \frac{\log\! M^+_i + \log\! M^-_i}{2} \, , \\
		s_i & = \frac{\log\! M^+_i - \log\! M^-_i}{2\dl} \, , \\
		n_i & = \left(i-\frac{1}{2}\right)\frac{1}{N_{sim} V} \, .
	\end{align}
	The $\pm$ superscripts refer to overdense and underdense separate universes. By fitting the above lists with splines, we obtained estimates of the functions $s(\log\! M)$, which is the mass shift, and $n(\log\! M)$, which is the halo mass function. The average Lagrangian bias above mass $M$ is then estimated
	\begin{align}
		b^L = -\frac{s(\log\! M)}{n(\log\! M)}\frac{dn(\log\! M)}{d\log\!M} \, .
	\end{align}
	Note, all biases reported and discussed below will be biases averaged above a mass threshold. 
	
	The halo finding program Rockstar \cite{Behroozi:2011ju} was used to extract catalogs of bound objects and their spherical overdense masses. Halo masses are calculated in Rockstar by finding the outermost particle position from the halo's center of mass at which the average spherical overdensity inside this radius is greater than some threshold,
	\begin{align}
		M_h & = M_p N \, , \\
		\frac{3NM_p}{4\pi r_N^3} > & \rho_{Th} > \frac{3(N+1)M_p}{4\pi r_{N+1}^3} \, .
	\end{align}
	The density threshold used here is the virialization threshold
	\begin{align}
		\frac{\rho_{Th}}{\bar{\rho}_m} & = \frac{18\pi^2 + 82 (\om(a)-1) - 39(\om(a)-1)^2}{\om(a)} \, .
	\end{align}
	Since Rockstar determines densities with respect to the local, separate universe comoving distances, the threshold has to be converted to these coordinates,
	\begin{align}
		\rho_{WTh} & = \rho_{Th}(1-\dl)\, .
	\end{align}
	That is, an overdense separate universe has an apparently lower density threshold to form collapsed objects.
	
	At small masses, the discreteness of this halo mass determination introduces an artificial spread in the values $s_i$. We modified this mass calculation, based on the work \cite{Li:2015jsz},  to include a contribution from particle $N+1$, assuming its mass to be uniformly distributed in the spherical shell between  $r_{N+1}$ and $r_N$,
	\begin{align}
		M_h & = M_p \lb N + \delta N \rb \, , \\
		\delta N & = \frac{M_p N - \rho_{Th} V_N}{\rho_{Th}(V_{N+1} - V_N^2) - M_p} \, .
	\end{align}	
	Here $V_N$ is the volume inside the radial position of the $N^{th}$ particle from the halo's center. Masses calculated this way are continuous, and produce less spread in the values of the mass shift at low mass. 
	
	We repeated this calculation of the Lagrangian bias, resampling with replacement over our sets of random initial conditions, and calculated the bootstrap variance. The biases at redshift $z=0$ are shown in FIG. \ref{fig:bs}, along with the ratio of the Lagrangian biases to the adiabatic ($\I  =0$) Lagrangian bias. We will refer to the latter quantity as the {\it relative  bias}.
	
	The scale dependence is clearly demonstrated within the bootstrap variances from the simulations, especially for the anticorrelated case. The recovery of the adiabatic bias below the Jeans scale is demonstrated for the $k/H_0=100$ mode. 

	The relative biases are consistent with being mass independent for the masses we are sensitive to $(\sim10^{13}$--$10^{15}\ \rm{M}_\odot)$. Small variations in the relative bias as a function of mass is due to the spline fitting. At masses above $10^{15}\ \rm{M}_\odot$, halos are rare and the spline fitting becomes very sensitive to the choice of knots, so the bias is not well determined there.

	As a test of our bias determinations, we compared them to measurements of the clustering bias from a simulation with box size $(1000\ \mathrm{Mpc})^3$, number of particles $(512)^3$, and a standard, global $\Lambda$CDM background cosmology with purely adiabatic perturbations. The clustering bias, $b_{{_C}}$, is measured from the matter-halo cross power spectrum on linear scales \cite{Dvali:2003ar, 2010PhRvD..82j3529D, 2014JCAP...08..056A},
	\begin{align}
		\frac{P_{mh}}{P_{mm}} \simeq b_c + b_2 k_S^2 + \mathcal{O}(k_S^4)\, .
	\end{align}
	Treating $b_2$ as a nuisance parameter, we fit the above function for value of $k_S$ up to $0.05\ \rm{Mpc}^{-1}$. We found the cluster bias to be in excellent agreement with our separate universe response biases for the adiabatic mode.
	
	\subsection{Linearity of responses}

	Our simulations involve two independent components of the long-wavelength modes: the curvature and the quintessence-sourced isocurvature. We fix the initial relative amplitudes of these as part of our simulation parameters. However, since both components of the long-wavelength mode stay perturbatively small, it should be the case that responses can be decomposed into two terms, one for each component.\footnote{It is always, of course, a working assumption of the separate universe approach that responses computed from simulations with a particular realization of the long-wavelength mode can be combined to produce the response for a different realization, e.g. for a larger or smaller amplitude $\delta_L$ or a $\delta_L$ with a different density profile. We nevertheless want to verify this assumption for the more extreme examples of different $\delta_L$ considered in this paper.}
	
	Suppose we have a small-scale observable $\mathcal{O}_W$, which responds to the long-wavelength mode,
	\begin{align}
		R_{\mathcal{O}} & = \frac{d\log \mathcal{O}_W}{d\dl} \, , \\
		&  = \frac{d\log \mathcal{O}_W}{d\dl^\R} \frac{\dl^\R}{\dl}  +\frac{d\log \mathcal{O}_W}{d\dl^\I} \frac{\dl^\I}{\dl} \,  .
	\end{align}
	In this way, we can take linear combinations of responses from simulations with purely adiabatic and purely isocurvature long-wavelength modes and obtain the response for any realization of $\dl$ with initial relative amplitude $\I/\R$.
	
	The evolution of the total long-wavelength matter perturbation with both curvature-sourced and isocurvature-sourced components can differ significantly from the purely isocurvature-sourced and purely curvature-sourced perturbations. The evolution even becomes nonmonotonic in the case with $\I/\R=-100$. In order to test the linearity of responses, an additional set of simulations was run with long-wavelength mode that was purely isocurvature-sourced, at wave number $k/H_0=10$. The power spectrum response and biases obtained by linearly combining simulation results with the pure isocurvature and pure curvature long-wavelength modes are shown in FIG. \ref{fig:lin}.
	
	Since the responses are estimated using finite difference derivatives with respect to the long-wavelength mode, we expect numerical errors on the order of $\delta_L\sim1\%$. In addition to this, we have cosmic variance from the limited set of realizations for the initial conditions. The difference in the power spectrum response obtained from linear combination and the total long-wavelength mode simulations is $<2\%$ even into the nonlinear regime, which is consistent with our expected numerical accuracy. 
	
	For the biases, the error is dominated by shot noise and stochasticity. At masses between $10^{13}$--$10^{15}\ \rm{M}_\odot$, the two estimates of the Lagrangian bias are within $1.5\ \sigma$, so linearly combining purely isocurvature-sourced and purely curvature-sourced simulation results gives a good estimate of the small-scale observable responses. As a result, our simulation responses can be rescaled and linearly combined to estimate what the responses would be in the presence of different long-wavelength modes with particular realizations of $\mathcal{I}$ and $\mathcal{R}$, and ultimately, different primordial power spectra and cross spectra for $\mathcal{I}$ and $\mathcal{R}$ as discussed in \ref{sec:SU}.
	
\section{Bias models}
\label{sec:biasmodels}

	The linearity of the biases allows us to identify two independent bias coefficients, $b_\R^L(a)$ and  $b_\I^L(a,k)$, for purely curvature-sourced and isocurvature-sourced matter perturbations respectively. In general, these have different mass dependence, so their ratio depends on both mass and wave number. For a given realization of $\I/\R$, the total relative bias is

	\begin{align}
		\frac{b^L(M,a;k)}{\blr(M,a)} = \frac{\dmr(a,k)}{\dm(a,k)}+ \frac{\bli(M,a;k)}{\blr(M,a)} \frac{\dmi(a,k)}{\dm(a,k)} \, .
	\end{align}
	To predict the total relative bias, we must model the relative bias for the purely isocurvature-sourced modes. This amounts to describing the scale dependence and time evolution of the pure isocurvature bias. We define the isocurvature relative bias,
	\begin{align}
		\beta^L(M,a;k) = \frac{\bli(M,a;k)}{\blr(M,a)} \, .
	\end{align}
	Different bias models correspond to different choices for the $k$-dependence and evolution of $\beta^L$.
	
	Below we consider two classes of models. The first class attempts to reproduce the pure isocurvature relative bias using the transfer functions $\tmr$ and $\tmi$. This fixes the scale dependence and evolution of the isocurvature bias up to an overall multiplicative factor, which we take to be  independent of scale and fit to the isocurvature dominated scales ($k/H_0 \leq 1.$). The transfer function models generically involve at least one free parameter. 
	
	The second class of models assumes that the bias is proportional to another small-scale observable's response to the long wavelength mode, such as the power spectrum response, or the response of the critical linear density for spherical collapse. The relative bias for this class of models is just the ratio between the pure isocurvature and pure curvature responses, so these models involve no free parameters.
	
	\subsection{Transfer function models}
	
	Models of halo bias that are local in time assume that the fluctuations in halo number density are fixed by the configuration of the matter density field at a single redshift. The evolution of halo abundance is then an initial value problem, with initial data given by the matter density field's configuration on a single time slice. This is, for instance, the point of view adopted in the excursion set theory approach to large scale structure. In this approach, the matter density field at early times is smoothed over a range of scales and compared to the critical density for spherical collapse linearly evolved back to the redshift of the initial data. From this point of view, it is natural to model the evolution and scale dependence of the halo bias using the transfer functions for the linear evolution of the different components of matter fluctuations. 
	
	For these models the Lagrangian biases have simple evolutions, inversely proportional to the transfer functions. We will consider two different possibilities for the evolution and scale dependence of $\beta(a,k)$. The first assumes passive evolution of halo abundance, so that the number density of halos at a given mass is conserved. Scale dependence in the bias then arises from scale dependence in the isocurvature transfer function at late times. The second model assumes that the individual biases are scale invariant at all times. Scale-dependent bias in this case arises from when the two bias terms are combined, through the scale dependence of the total matter perturbation.
	
	The models we consider are based on the matter transfer functions $\tmr$ and $\tmi$, which is motivated by the linearity of the Lagrangian response bias. Another approach would be to include a bias term for the quintessence density contrast ($b_Q \delta_Q$), combined with a bias term for either the full matter perturbation, or just the curvature sourced part. This is similar to including a neutrino bias term for the case of scale dependent growth arising from massive neutrinos. This approach was found not to reproduce biases from separate universe simulations \cite{Chiang:2017vuk}. Similarly, we find that the scale dependence from the quintessence transfer function, when combined with the matter transfer functions, does not reproduce the scale dependence of the simulation biases.  Fits from these models have $\chi^2 \sim 10^3$ (per degree of freedom),  so we do not consider them below.

	\subsubsection{Passive evolution model}
		
	\begin{figure*}
		\includegraphics[width=\linewidth]{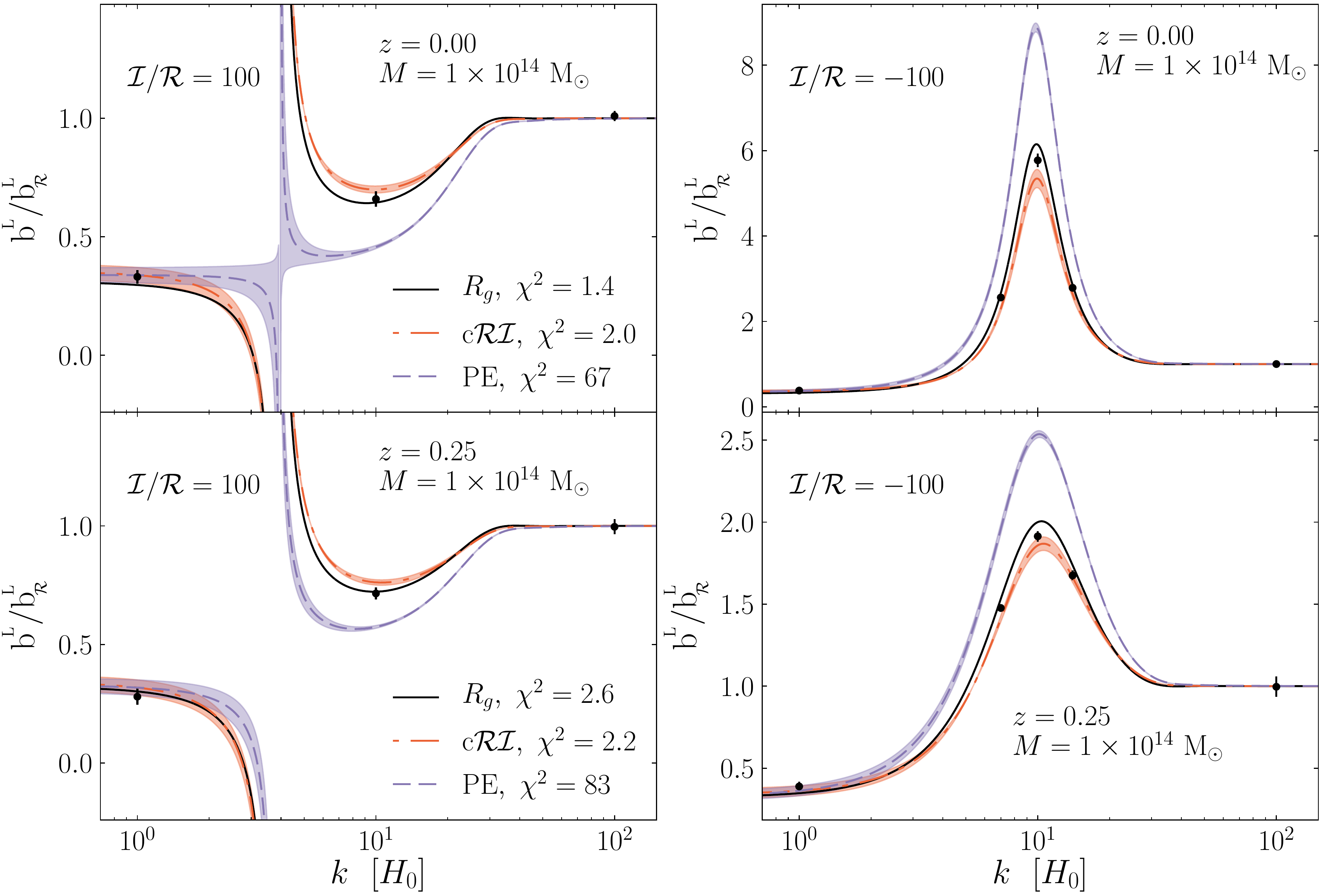}
		\caption{Model comparisons for the Lagrangian relative bias, at redshift $z=0$ on top and mass $z=0.25$ on the bottom. The data points in black are the simulation results, and their error bars show the bootstrap variance taken at mass $M=10^{14}\ \rm{M}_\odot$. We choose to show $M=10^{14}\ \rm{M}_\odot$ since it is relatively well constrained but since the relative bias is consistent with being mass independent (FIG. \ref{fig:bs}) model comparisons at other masses will look similar. The $\chi^2$ values shown for each model are per degree of freedom, jointly fitting both $\I/\R=\pm100$ simulation biases. Note that for $k\ll k_{Jeans}$ the ratio $b^L/b^L_{\R}$ is nearly the same for both values of $\I/\R = \pm 100$, this is because the isocurvature-generated matter fluctuations dominate $\delta_m$ at low-$k$. }	
		\label{fig:bmods}
	\end{figure*}
	
	A simple example of bias evolution, based on work by Hui and Parfrey \cite{Hui:2007zh, Parfrey:2010uy}, assumes that the number density contrast of halos is conserved,
	\begin{align}
		\delta_h(M,a_i) = \delta_h(M,a) \, .
	\end{align}
	This passive evolution model (PE in what follows) is accurate so long as merger events between different halos are sufficiently rare. The evolution of the individual biases is given by
	\begin{align}
		\blr(M,a) & = \blr(M,a_i) \frac{\tmr(a_i,k)}{\tmr(a,k)} \, , \\  
		\bli(M,a;k) & =\bli(M,a_i) \frac{\tmi(a_i)}{\tmi(a,k)} \, .
	\end{align}
	Each of these individual bias terms are independent of the realization of the long-wavelength mode. The scale dependence from the initial condition in $\tmr$ cancels taking the ratio. While the initial condition for $\tmi$ does not depend on $k$, its subsequent evolution after horizon crossing does, and this is where the scale dependent bias originates in this model.

	The relative bias for pure isocurvature is
	\begin{align}
		\beta^{L}_{PE}(M,a;k) & = \beta^{L}_{PE}(M,a_i) \frac{\tmr(a,k)\tmi(a_i)}{\tmr(a_i,k)\tmi(a,k)} \, .
	\end{align}
	The free parameter $ \beta^L_{PE}(M, a_i)\equiv \bli(M,a_i)/\blr(M,a_i)$, can be fit by requiring this relative bias to match the simulation results on isocurvature dominated scales.

	The total relative bias can be written
	\begin{align}
		\label{eq:LPH}
		\frac{b^{L}(M,a;k)}{\blr(M,a)}   =  \lb 1 + \beta_{PE}(M,a_i)\frac{\dmi(a_i,k)}{\dmr(a_i,k)} \rb	\frac{\dmr(a)}{\dm(a,k)} \, .
	\end{align}
    Since the initial transfer function $\tmr(a_i,k)$ grows as $k^2$, the correct small-scale relative bias  $b^E/b^E_\R=1$ is recovered in the curvature dominated regime ($k\rightarrow\infty$).
	
	The model fits at mass $M=10^{14}\ \rm{M}_\odot$ are shown in FIG. \ref{fig:bmods} at redshifts $z=0$ and $z=0.25$. The model does not reproduce the scale dependence of our simulation results. Forcing the model to agree with the isocurvature dominated results causes the relative bias to be dramatically over estimated for $\I/\R=-100$. For $\I/\R=100$, the relative bias at the Jeans scale is underestimated. Similar versions of this model were also shown to give poor reproduction of simulation results for adiabatic quintessence \cite{Chiang:2016vxa} and massive neutrinos \cite{Chiang:2017vuk}.

	\subsubsection{Constant $\R$-$\I$ bias model}
	
    Suppose that, instead of taking the evolution of $\bli$ to be determined by the assumption of halo number density contrast conservation, we instead take it to be scale independent at all times. Then the total relative bias is given by
	\begin{align}
		\frac{ b^{L}(M,a;k)}{ \blr(M,a)} = \frac{\dmr(a,k)}{\dm(a,k)} + \beta_{\R\I}(M,a) \frac{ \dmi(a,k)}{ \dm(a,k)} \, .
	\end{align}
	We can again fit the parameter $\beta_{\R\I}^{L}(M, a)$ to match the isocurvature dominated scales at each redshift. We call this model the constant $\R$-$\I$ bias model (c$\R\I$). Although the individual components of the bias are assumed to be independent of scale, the total bias does depend on scale through the isocurvature transfer function. The component biases each depend on mass and redshift, and there is nothing requiring the relative bias to be independent of mass for this model. 

	Fits for this model are have $\chi^2\simeq2$ per degree of freedom. For $\I/\R=100$, the Jeans scale relative bias is slightly over estimated, whereas the $\I/\R=-100$ biases are underestimated.
	
	Overall, the transfer function models appear to capture the qualitative features of the relative biases. The scale dependence seems to be dominated by the shape of the function $\delta_m(a,k)^{-1}$, which is what one expects for passive halo evolution. Quantitatively, the c$\R\I$ model gives a much improved reproduction of the simulation biases compared with the PE model. The quality of these model fits is not strongly dependent on redshift. The relative success of the c$\R\I$ model indicates that the scale dependence of the isocurvature relative bias is fairly weak. 
	
	\subsection{Response models}

	If the number density of collapsed objects cannot be determined from the configuration of the density field at a single time, but instead relies on the cumulative growth history of the matter density, then the bias is nonlocal in time.  In this case, we can try to characterize the local mass function as depending on some other small scale observable, which, in the separate universe context depends on the long-wavelength modes that are present in a given region,
	\begin{align}
		n_W(M,a;k) = n(M,a; \mathcal{O}_W(a,k)) \, .
	\end{align}
	The response bias is then proportional to the growth response of $\mathcal{O}_W$,
	\begin{align}
		\label{eq:bLresponse}
		b^L(M,a;k) = \frac{d\log n}{d\log\mathcal{O}}(M,a) R_{\mathcal{O}}(a,k) \, .
	\end{align}
	The first factor on the right side of the above equation is evaluated in the global universe, so it is only a function of mass, time, and the global value of the quantity $\mathcal{O}$. That is, this factor is independent of the long-wavelength mode. Then the relative bias is
	\begin{align}
		\beta^L_{\mathcal{O}}(a,k) = \frac{R^\I_{\mathcal{O}}(a,k) }{R^\R_{\mathcal{O}}(a) } \, .
	\end{align} 
	Notice that for this class of models, the relative bias is predicted to be independent of mass\footnote{If Eq.~(\ref{eq:bLresponse}) were evaluated at the object formation time, rather than the observation time, then the relative bias would be mass-dependent since objects of different masses typically form at different times and are therefore sensitive to $R_{\mathcal{O}}(a,k)$ at different epochs.}. Also, unlike the transfer function models, these models have no free parameters. The relative biases are completely determined by linear perturbation theory and the separate universe formalism.
	
	We considered response models based on two possible choices for $\mathcal{O}$: the local power spectrum, and the local critical density for spherical collapse \cite{LoVerde:2014pxa}. These were shown in \cite{Chiang:2016vxa, Chiang:2017vuk, Chiang:2018laa} to perform better than bias models based on transfer functions. We found that the results of the spherical collapse model match the results of the power spectrum response model, so we consider only the latter model in what follows.
	
	Assume the local mass function in a region depends on the local power spectrum in a universal way, then the pure isocurvature relative bias is given
	\begin{align}
		\beta^{L}_{R_P}(a,k) & = \frac{ R^\I_P(a,k)}{ R_P^{\R}(a)} \, .
	\end{align}

	\begin{figure}
		\includegraphics[width=\linewidth]{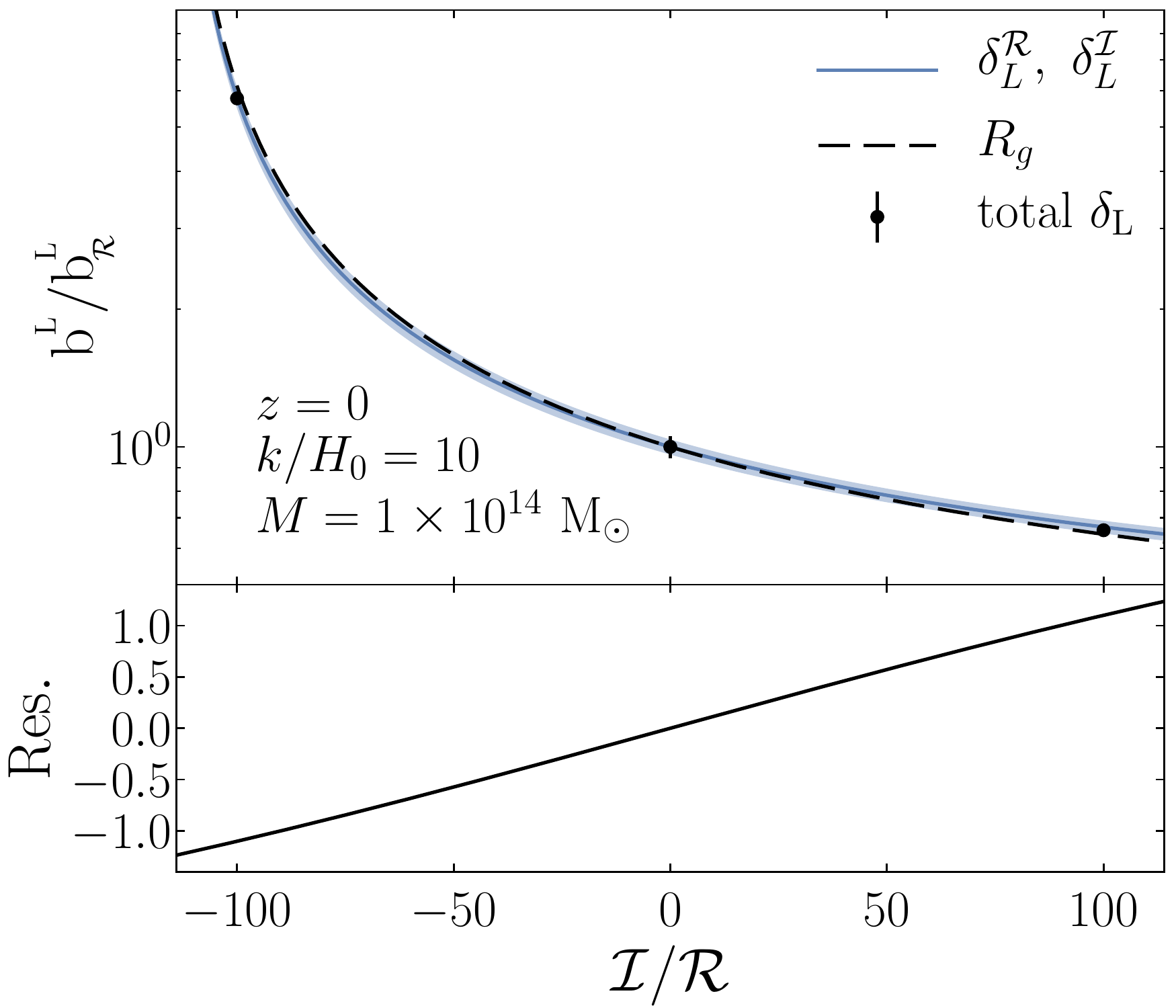}
		\caption{The solid blue line shows the linear combination of purely isocurvature and purely curvature simulation biases, with the shaded area showing the bootstrap variance. The data points, from left to right are from simulations with $\I/\R=-100,\ 0,$ and $100$. The dashed line is the prediction from the power spectrum response bias model. The bottom panel shows the residuals between the linearly combined relative biases and the power spectrum response model. }	
		\label{fig:bk10RI}
	\end{figure}
	
	This model is compared in FIG. \ref{fig:bmods} with relative biases from the simulations at $M=10^{14}\ \rm{M}_\odot$ for redshifts $z=0$ and $z=0.25$. We obtained $\chi^2 = 1.4 $ per degree of freedom at the later redshift, and $\chi^2 = 2.6$ at the earlier redshift. It slightly underpredicts the Jeans scale bias in the $\I/\R=100$ simulations, and overpredicts the $\I/\R=-100$ biases.

	At smaller values of $\I/\R$, the power spectrum response model does much better. The linearly combined pure isocurvature and pure curvature relative biases are shown in FIG. \ref{fig:bk10RI}, for $k/H_0=10$. Agreement between the linear combination bias and the $R_P$ model is good for $|\I/\R| < 100$.

\section{Observables}
\label{sec:observables}	

	\begin{figure*}
		\includegraphics[width=\linewidth]{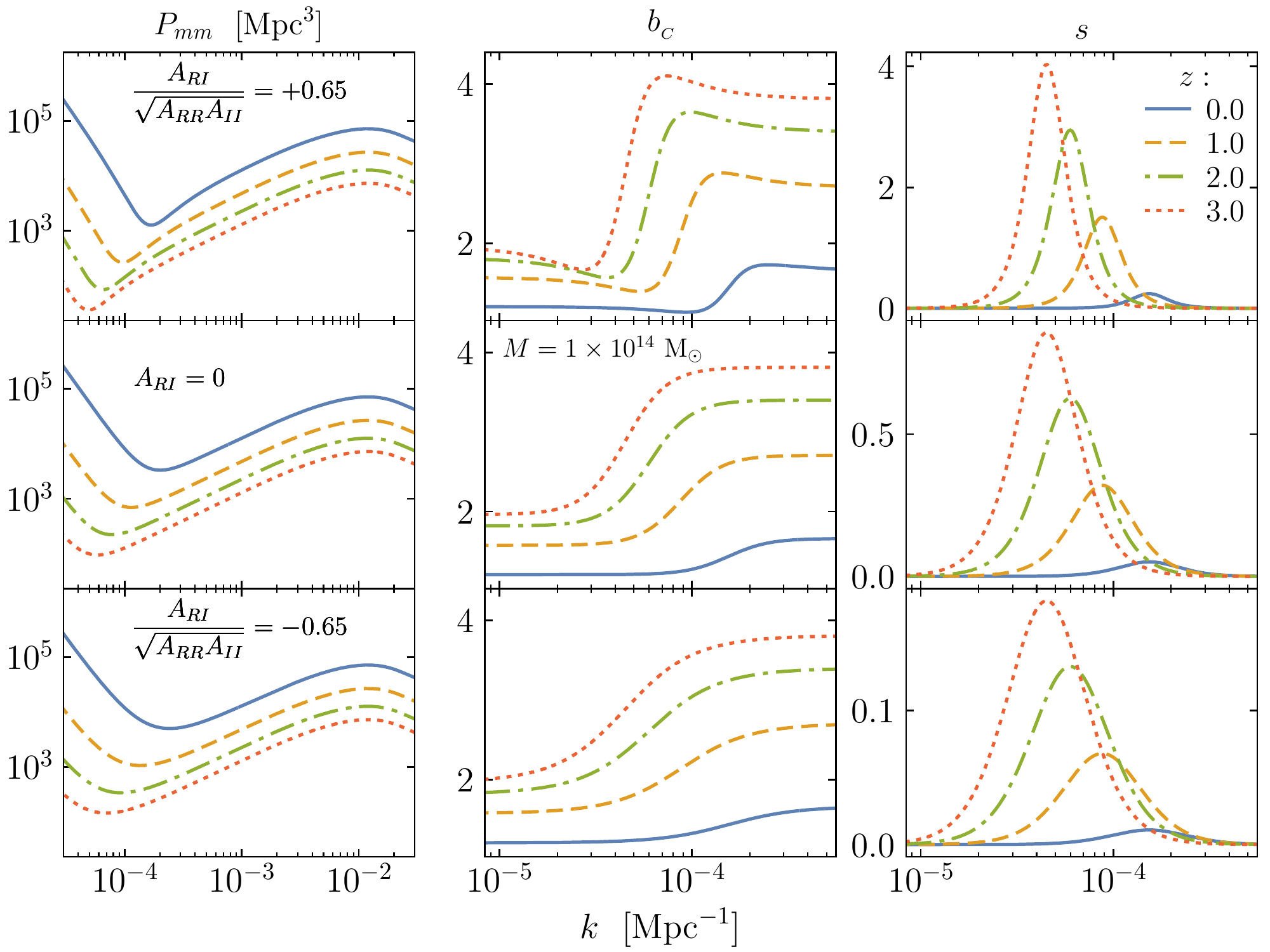}	
		\caption{From left to right shows the matter power spectrum, the clustering bias, and the stochasticity at redshifts from $z=0$--$3$. In this case, we took $A_{II}=A_{RR}$. The clustering bias and stochasticity are averaged for halos with mass greater than $M=10^{14}\ \rm{M}_{\odot}$. Since the relative bias is consistent with being independent of mass over a wide range of masses, only the adiabatic bias is required to determine $b_c$ and $s$ at other masses. Top to bottom shows the cases of isocurvature and curvature that are correlated, uncorrelated and anticorrelated. }
		\label{fig:obs}
	\end{figure*}
	
	In real surveys or in large box simulations, the regions covered can be orders of magnitude above the quintessence Jeans scale considered here. The long-wavelength mode in these cases can also be sampled over. The effects of averaging over $\delta_L$ to obtain the squeeze limit bispectrum were already shown in FIG. \ref{fig:Rs}, assuming scale invariant auto and cross primordial power spectra. Here will consider additional observables, including the large-scale matter power spectrum, the clustering bias, and the stochasticity. 
		
	Clustering bias and stochasticity are defined by taking correlations between the long-wavelength matter perturbations and the halo number of density fluctuations (e.g. \cite{Matsubara:1999qq, Hui:2007zh,Smith:2010gx}). Writing separate bias terms for the curvature and isocurvature modes
	\begin{align}
		\delta_h  = b_\R \dl^R + b_\I\dl^\I \, ,
	\end{align}
	where $\delta_h$ is the density contrast for the number halos and we have suppressed all arguments, including the $k$-dependence in $b_\I$, $\dl^R $, and $\dl^\I$, for notational clarity. We then have the following three power spectra for matter perturbations and halos:
	\begin{align}
		P_{mm} =\ & (\tmr)^2P_{\R\R} + 2\tmr\tmi P_{\R\I} + (\tmi)^2P_{\I\I} \, , \\ 
		\nonumber \\
		P_{mh} =\ & b_\R (\tmr)^2 P_{\R\R} + \lb b_\R + b_\I \rb \tmr\tmi P_{\R\I} \\ & + b_\I(\tmi)^2 P_{\I\I} \, ,  \nonumber \\
		\nonumber \\
		P_{hh} =\ & b_\R^2 (\tmr)^2 P_{\R\R} + 2\ b_\R b_\I \tmr\tmi P_{\R\I} \\ & + b_\I^2 (\tmi)^2 P_{\I\I} \, .  \nonumber
	\end{align}
	Note that for $b_\R\neq b_\I$ the above set of equations is nondegenerate so that equations can be inverted to solve for the individual contributions to the matter power spectrum sourced by $\I$ and $\R$, e.g. $P_{mm}^{\R\R}$, $P_{mm}^{\I\R}$, and $P_{mm}^{\I\I}$, from the observed $P_{mm}$, $P_{mh}$ and $P_{hh}$. The clustering bias is defined as
	\begin{align}
		b_{{}_C} = \frac{P_{mh}}{P_{mm}} \, .
	\end{align}
	The stochasticity is defined
	\begin{align}
		s = \frac{P_{hh}P_{mm} - \lb P_{mh}\rb^2}{\lb P_{mm}\rb^2} \, .
	\end{align}
	Here we will assume that all three primordial power spectra are scale invariant, and we take the amplitude of isocurvature fluctuations to equal the amplitude of scalar curvature fluctuations. Finally, we consider cases where the cross power spectrum is $65\%$ correlated or anticorrelated, and the case where it is totally uncorrelated.	
	
	Since we take the isocurvature amplitude to equal the curvature amplitude, we are well within the regime where the parameter-free power spectrum response model adequately describes the bias.
	\begin{align}
		b_{X} = 1 + \lb \frac{d\log n}{d\log P} \rb R_P^{X} \, ,
	\end{align}
	where $X$ can be either $\R$ or $I$. The biases in this section are all Eulerian ($b^E = 1+b^L $), although we will omit the superscript $E$ that indicate this. 
	
	Using the biases from adiabatic mode simulations, the mass and redshift dependent function can be fit and used to calculate the power spectrum response model biases at large-scales in the presence of isocurvature. With the bias determined at five redshifts, we fit this for mass $M=1\times 10^{14}\ \rm{M}_{\odot}$ as a quadratic function of scale factor.   

	These observables are plotted in FIG. \ref{fig:obs} for redshifts ranging from $z=0$--$3$. The most significant feature predicted by this model is the growth of power at very large scales with $k<10^{-4}\ \rm{Mpc}^{-1}$. This is the isocurvature curvature dominated regime, above the quintessence Jeans scale. Current observations do not yet extend to scales with $k\lesssim10^{-3}\ \rm{Mpc}^{-1}$ \cite{Tegmark:2002cy, 2012ApJ...749...90H}. 
	
	The growth of the power spectrum at large, isocurvature dominated scales is due to that fact that $\tmi/\tmr \sim k^{-4}$.  As long as $P_{\I\I}/P_{\R\R}$ does not scale as a higher power of $k$ than $k^4$, $P^{\I\I}_{mm}$ dominates the power spectrum on scales far above the Jeans scale ($k \ll k_{Jeans}\simeq2\times10^{-3}\  \rm{Mpc}^{-1}$), and the power spectrum will increase with decreasing wave number. Below the Jeans scale, $P^{\R\R}_{mm}$ dominates, so the typical $\Lambda$CDM matter power spectrum with adiabatic initial conditions is recovered on these scales. 
	
	Near the Jeans scale, the shape and evolution of the matter power spectrum depends on the curvature-isocurvature cross-correlation term $P^{\R\I}_{mm}$, as well as the ratio $P_{\I\I}/P_{\R\R}$. For positive cross-correlation, the total matter power spectrum has excess power compared with $P^{\R\R}_{mm}$ at the Jeans scale due to the peak in the isocurvature transfer function. The power spectrum drops below $P^{\R\R}_{mm}$ at larger scales due to the change in sign of the isocurvature transfer function. In the limit of $100\%$ correlation, there is a wave number above the Jeans scale at which the power spectrum vanishes. The total power spectrum increases as $P_{\I\I}$ starts to dominate near and above the horizon scale. Similarly, in the anticorrelated case, the total matter power spectrum at the Jeans scale is deficient compared with $P^{\R\R}_{mm}$. For the parameter choices in FIG. \ref{fig:obs}, the deviation between $P_{mm}$ and $P^{\R\R}_{mm}$ is about $1\%$ at the Jeans scale. For parameters comparable to the single realizations of our separate universe simulations, $A_{\I\I} = 10^4 A_{\R\R}$, the Jeans scale matter power spectrum differs from $P_{\R\R}$ by about $300\%$. However, the deviation is less than $1\%$ at scales below $k\gtrsim8\times10^{-3}\ \mathrm{Mpc^{-1}}$.
	
	There is also clear scale dependence of the clustering bias above the Jeans scale. This scale dependence is sensitive to the cross-correlation between curvature and isocurvature fluctuations. The stochasticity is peaked at a scale above the Jeans scale. The location of this peak depends on the ratio of the curvature and isocurvature amplitudes. For larger isocurvature amplitudes, this scale approaches the quintessence Jeans scale. The height and width of the peaks in the stochasticity are sensitive to the cross-correlation. Over time, both the clustering bias and stochasticity evolve to approach unity, so the scale dependence is most prominent at earlier redshifts. However, the scale dependence shows up at larger scales at earlier times.
	
	For completeness we note that the presence an isocurvature component in the matter perturbations will also have an effect on redshift space distortions. An object, such as a halo, has redshift space coordinate $\vec{s}$ defined by
	\begin{align}
		\vec{s} = \left(r + \hat{r}\cdot\vec{v}\right)\hat{r}\, ,
	\end{align}  
	where $r$ is the radial distance to the object, $\vec{v}$ is the peculiar velocity of the object, and $\hat{r}$ is the line-of-site direction. Here we will consider the form of the redshift space matter power spectrum derived by Kaiser \cite{Kaiser:1987qv}. The Jacobian of the transformation from real space to redshift space, and  the linear matter continuity equation can be used to express the Fourier modes of the density contrast for halo numbers in redshift space,
	\begin{align}
		\delta^{s}_h(\vec{k}, a) = \delta^{r}_h(\vec{k}, a) + \mu(\vec{k})^2\delta^{r}_m(\vec{k},a)'\, .
	\end{align} 
	The coefficient $\mu(\vec{k})$ is the cosine of the angle between the wave vector and the line of sight. The superscripts $s$ and $r$ denote the density contrast in redshift space and real space respectively. Substituting the full matter density contrast with transfer functions and the halo biases, the halo power spectrum in redshift space, $P^{s}_{mm}$, is
	\begin{align}
		P_{hh}^{s}  = & \left(b_\R^2 P_{mm}^{\R\R}+ 2b_\R b_\I P_{mm}^{\R\I}+ b_\I^2 P_{mm}^{\I\I}\right) \\ 
		& + 2\mu^2b_\R \left(f_\R P_{mm}^{\R\R} + f_\I P_{mm}^{\R\I} \right) \nonumber \\
		& + 2\mu^2b_\I\left( f_I P_{mm}^{\I\I} + f_\R P_{mm}^{\R\I} \right) \nonumber \\
		&+  \mu^4\left(f_\R^2 P_{mm}^{\R\R}+ 2f_\R f_\I P_{mm}^{\R\I}+ f_\I^2P_{mm}^{\I\I}\right)\, , \nonumber
	\end{align}
	where we have defined $f_\R \equiv d\log T_m^\R/d\log a$, $f_\I \equiv d\log T_m^\I/d\log a$. Unlike in the case of purely adiabatic fluctuations, taking the ratio of the $\mu^4$ terms with the $\mu^2$ terms does not isolate the growth rate and bias from the primordial spectra. On the other hand, at each order in $\mu^2$ the terms in the redshift space power spectrum contain distinct weightings of $P_{mm}^{\R\R}$, $P_{mm}^{\I\R}$, and $P_{mm}^{\I\I}$ so that, even if $b_\R =b_\I$, the isocurvature and curvature power spectra could be solved for in terms of $f_\R$, $f_\I$, $b_\R$, and $b_\I$.
	
	Since the quintessence isocurvature perturbations are only important at large scales and at late times, it is difficult to constrain the amplitude of the isocurvature auto power spectrum and the correlation between isocurvature and curvature fluctuations. For instance, an analysis of the 7-year WMAP data allowing for dark energy isocurvature leaves the dark energy sound speed and primordial isocurvature power spectra and cross-spectra virtually unconstrained \cite{Liu:2010ba}. A full analysis of this model with current CMB temperature and polarization data, along with large-scale structure data is needed to determine allowed ranges of the parameters $c_Q$, $P_{\I\I}$, and $P_{\I\R}$.
		
\section{Conclusion}	
	
	Quintessence isocurvature perturbations provide a scenario in which large-scale structure formation can be studied in a context that is more general than the canonical $\Lambda$CDM model. The scalar field dark energy perturbations introduce both scale dependent growth, through the presence of a quintessence Jeans scale, and also initial condition dependent growth history. As shown in FIG. \ref{fig:d}, large-scale matter perturbations can have dramatically different, and even nonmonotonic evolution in this context, depending on scale and initial conditions.
	
	Using separate universe techniques and N-body simulations, we have studied the responses of small-scale observables such as the local power spectrum (FIG. \ref{fig:dP}) and halo mass function  (FIG. \ref{fig:bs}) to the presence of long-wavelength perturbations. Our methods have been validated by comparing the results of the power spectrum response to 1-loop calculations from perturbation theory and by comparing the response biases for the adiabatic case to the clustering bias in larger volume simulations. We have also verified, in FIG. \ref{fig:lin}, that the individual responses to long-wavelength curvature and isocurvature-sourced perturbations can be linearly combined to study the effects of varying the initial conditions. That is, the separate universe simulations can be used to make predictions for any initial large-scale curvature and isocurvature power spectra. 
	
	The linearity of small-scale observable responses guarantees that the evolution of the total bias depends on both scale and initial conditions. While scale-dependent bias has previously been found, originating from scale dependent growth due to neutrinos and quintessence without isocurvature, this is the first instance of bias evolution that depends on initial conditions of the long-wavelength matter modes. 
	
	In section \ref{sec:biasmodels}, we tested a number of simple bias evolution models against our simulation results. These models break down into two classes: transfer function models and response models. While all of the models roughly capture the dependence on scale and initial conditions of the bias evolution, FIG. \ref{fig:bmods} demonstrates that two of the models considered reproduce the scale dependent features in the bias. One is the the transfer function model that assumed scale invariant bias factors for both the curvature-sourced and isocurvature-sourced components of the matter perturbations (the c$\R\I$ model). The success of this bias model is a consequence of the linearity of the response observables, and the weak scale dependence of the purely isocurvature-sourced halo bias. The other viable bias model was based on the power spectrum response ($R_g$). This model outperformed the c$\R\I$ model at low redshift, but was worse by redshift $z=0.25$. On the other hand, the response model is quite accurate for more modest amplitudes of isocurvature perturbations (see. FIG.\ref{fig:bk10RI}). Since this model has no free parameters, it can be used to make predictions without running additional simulations.  The third model we considered, based on an assumption of passive halo evolution (PE), gave poor reproduction of the simulation results at all masses and redshifts.
	
	The dependence of small-scale observables on growth history, through the wave number and initial conditions of a long-wavelength mode, indicates that structure formation is nonlocal in time. That is, it is not enough to know the statistics of the density field at a single redshift in order to predict the statistics of halos at another redshift. In this work we have provided a demonstration of history dependent structure formation in a model where the effect can be made arbitrarily large, in the sense that the ratio of isocurvature to curvature can be large. In principle, the relative strength of isocurvature fluctuations to curvature, as well as the quintessence sound speed can be constrained by large-scale structure observations and CMB measurements. 
	
\acknowledgements

	DJ and ML are grateful for helpful correspondence and conversations with Niayesh Afshordi, Chi-Ting Chiang, Wayne Hu, Yin Li,  and Ravi Sheth. Results in this paper were obtained using the high-performance computing system at the Institute for Advanced Computational Science at Stony Brook University. DJ is supported by Grants No. NSF PHY-1620628 and DOE DE-SC0017848. ML is supported by Grant No. DOE DE-SC0017848.
	
\bibliography{q_iso_su.bib}

\begin{thebibliography}{53}%
\makeatletter
\providecommand \@ifxundefined [1]{%
 \@ifx{#1\undefined}
}%
\providecommand \@ifnum [1]{%
 \ifnum #1\expandafter \@firstoftwo
 \else \expandafter \@secondoftwo
 \fi
}%
\providecommand \@ifx [1]{%
 \ifx #1\expandafter \@firstoftwo
 \else \expandafter \@secondoftwo
 \fi
}%
\providecommand \natexlab [1]{#1}%
\providecommand \enquote  [1]{``#1''}%
\providecommand \bibnamefont  [1]{#1}%
\providecommand \bibfnamefont [1]{#1}%
\providecommand \citenamefont [1]{#1}%
\providecommand \href@noop [0]{\@secondoftwo}%
\providecommand \href [0]{\begingroup \@sanitize@url \@href}%
\providecommand \@href[1]{\@@startlink{#1}\@@href}%
\providecommand \@@href[1]{\endgroup#1\@@endlink}%
\providecommand \@sanitize@url [0]{\catcode `\\12\catcode `\$12\catcode
  `\&12\catcode `\#12\catcode `\^12\catcode `\_12\catcode `\%12\relax}%
\providecommand \@@startlink[1]{}%
\providecommand \@@endlink[0]{}%
\providecommand \url  [0]{\begingroup\@sanitize@url \@url }%
\providecommand \@url [1]{\endgroup\@href {#1}{\urlprefix }}%
\providecommand \urlprefix  [0]{URL }%
\providecommand \Eprint [0]{\href }%
\providecommand \doibase [0]{http://dx.doi.org/}%
\providecommand \selectlanguage [0]{\@gobble}%
\providecommand \bibinfo  [0]{\@secondoftwo}%
\providecommand \bibfield  [0]{\@secondoftwo}%
\providecommand \translation [1]{[#1]}%
\providecommand \BibitemOpen [0]{}%
\providecommand \bibitemStop [0]{}%
\providecommand \bibitemNoStop [0]{.\EOS\space}%
\providecommand \EOS [0]{\spacefactor3000\relax}%
\providecommand \BibitemShut  [1]{\csname bibitem#1\endcsname}%
\let\auto@bib@innerbib\@empty
\bibitem [{\citenamefont {Maldacena}(2003)}]{Maldacena:2002vr}%
  \BibitemOpen
  \bibfield  {author} {\bibinfo {author} {\bibfnamefont {J.~M.}\ \bibnamefont
  {Maldacena}},\ }\href {\doibase 10.1088/1126-6708/2003/05/013} {\bibfield
  {journal} {\bibinfo  {journal} {JHEP}\ }\textbf {\bibinfo {volume} {05}},\
  \bibinfo {pages} {013} (\bibinfo {year} {2003})},\ \Eprint
  {http://arxiv.org/abs/astro-ph/0210603} {arXiv:astro-ph/0210603 [astro-ph]}
  \BibitemShut {NoStop}%
\bibitem [{\citenamefont {Creminelli}\ and\ \citenamefont
  {Zaldarriaga}(2004)}]{Creminelli:2004yq}%
  \BibitemOpen
  \bibfield  {author} {\bibinfo {author} {\bibfnamefont {P.}~\bibnamefont
  {Creminelli}}\ and\ \bibinfo {author} {\bibfnamefont {M.}~\bibnamefont
  {Zaldarriaga}},\ }\href {\doibase 10.1088/1475-7516/2004/10/006} {\bibfield
  {journal} {\bibinfo  {journal} {JCAP}\ }\textbf {\bibinfo {volume} {0410}},\
  \bibinfo {pages} {006} (\bibinfo {year} {2004})},\ \Eprint
  {http://arxiv.org/abs/astro-ph/0407059} {arXiv:astro-ph/0407059 [astro-ph]}
  \BibitemShut {NoStop}%
\bibitem [{\citenamefont {Grin}\ \emph {et~al.}(2011)\citenamefont {Grin},
  \citenamefont {Dore},\ and\ \citenamefont {Kamionkowski}}]{Grin:2011tf}%
  \BibitemOpen
  \bibfield  {author} {\bibinfo {author} {\bibfnamefont {D.}~\bibnamefont
  {Grin}}, \bibinfo {author} {\bibfnamefont {O.}~\bibnamefont {Dore}}, \ and\
  \bibinfo {author} {\bibfnamefont {M.}~\bibnamefont {Kamionkowski}},\ }\href
  {\doibase 10.1103/PhysRevD.84.123003} {\bibfield  {journal} {\bibinfo
  {journal} {Phys. Rev.}\ }\textbf {\bibinfo {volume} {D84}},\ \bibinfo {pages}
  {123003} (\bibinfo {year} {2011})},\ \Eprint {http://arxiv.org/abs/1107.5047}
  {arXiv:1107.5047 [astro-ph.CO]} \BibitemShut {NoStop}%
\bibitem [{\citenamefont {McDonald}(2003)}]{McDonald:2001fe}%
  \BibitemOpen
  \bibfield  {author} {\bibinfo {author} {\bibfnamefont {P.}~\bibnamefont
  {McDonald}},\ }\href {\doibase 10.1086/345945} {\bibfield  {journal}
  {\bibinfo  {journal} {The Astrophysical Journal}\ }\textbf {\bibinfo {volume}
  {585}},\ \bibinfo {pages} {34} (\bibinfo {year} {2003})}\BibitemShut
  {NoStop}%
\bibitem [{\citenamefont {Sirko}(2005)}]{Sirko:2005uz}%
  \BibitemOpen
  \bibfield  {author} {\bibinfo {author} {\bibfnamefont {E.}~\bibnamefont
  {Sirko}},\ }\href {\doibase 10.1086/497090} {\bibfield  {journal} {\bibinfo
  {journal} {Astrophys. J.}\ }\textbf {\bibinfo {volume} {634}},\ \bibinfo
  {pages} {728} (\bibinfo {year} {2005})},\ \Eprint
  {http://arxiv.org/abs/astro-ph/0503106} {arXiv:astro-ph/0503106 [astro-ph]}
  \BibitemShut {NoStop}%
\bibitem [{\citenamefont {Wagner}\ \emph {et~al.}(2015)\citenamefont {Wagner},
  \citenamefont {Schmidt}, \citenamefont {Chiang},\ and\ \citenamefont
  {Komatsu}}]{Wagner:2014aka}%
  \BibitemOpen
  \bibfield  {author} {\bibinfo {author} {\bibfnamefont {C.}~\bibnamefont
  {Wagner}}, \bibinfo {author} {\bibfnamefont {F.}~\bibnamefont {Schmidt}},
  \bibinfo {author} {\bibfnamefont {C.-T.}\ \bibnamefont {Chiang}}, \ and\
  \bibinfo {author} {\bibfnamefont {E.}~\bibnamefont {Komatsu}},\ }\href
  {\doibase 10.1093/mnrasl/slu187} {\bibfield  {journal} {\bibinfo  {journal}
  {Mon. Not. Roy. Astron. Soc.}\ }\textbf {\bibinfo {volume} {448}},\ \bibinfo
  {pages} {L11} (\bibinfo {year} {2015})},\ \Eprint
  {http://arxiv.org/abs/1409.6294} {arXiv:1409.6294 [astro-ph.CO]} \BibitemShut
  {NoStop}%
\bibitem [{\citenamefont {Dai}\ \emph {et~al.}(2015)\citenamefont {Dai},
  \citenamefont {Pajer},\ and\ \citenamefont {Schmidt}}]{Dai:2015jaa}%
  \BibitemOpen
  \bibfield  {author} {\bibinfo {author} {\bibfnamefont {L.}~\bibnamefont
  {Dai}}, \bibinfo {author} {\bibfnamefont {E.}~\bibnamefont {Pajer}}, \ and\
  \bibinfo {author} {\bibfnamefont {F.}~\bibnamefont {Schmidt}},\ }\href
  {\doibase 10.1088/1475-7516/2015/10/059} {\bibfield  {journal} {\bibinfo
  {journal} {JCAP}\ }\textbf {\bibinfo {volume} {1510}},\ \bibinfo {pages}
  {059} (\bibinfo {year} {2015})},\ \Eprint {http://arxiv.org/abs/1504.00351}
  {arXiv:1504.00351 [astro-ph.CO]} \BibitemShut {NoStop}%
\bibitem [{\citenamefont {Hu}\ \emph {et~al.}(2016)\citenamefont {Hu},
  \citenamefont {Chiang}, \citenamefont {Li},\ and\ \citenamefont
  {LoVerde}}]{Hu:2016ssz}%
  \BibitemOpen
  \bibfield  {author} {\bibinfo {author} {\bibfnamefont {W.}~\bibnamefont
  {Hu}}, \bibinfo {author} {\bibfnamefont {C.-T.}\ \bibnamefont {Chiang}},
  \bibinfo {author} {\bibfnamefont {Y.}~\bibnamefont {Li}}, \ and\ \bibinfo
  {author} {\bibfnamefont {M.}~\bibnamefont {LoVerde}},\ }\href {\doibase
  10.1103/PhysRevD.94.023002} {\bibfield  {journal} {\bibinfo  {journal} {Phys.
  Rev.}\ }\textbf {\bibinfo {volume} {D94}},\ \bibinfo {pages} {023002}
  (\bibinfo {year} {2016})},\ \Eprint {http://arxiv.org/abs/1605.01412}
  {arXiv:1605.01412 [astro-ph.CO]} \BibitemShut {NoStop}%
\bibitem [{\citenamefont {Li}\ \emph {et~al.}(2014{\natexlab{a}})\citenamefont
  {Li}, \citenamefont {Hu},\ and\ \citenamefont {Takada}}]{Li:2014sga}%
  \BibitemOpen
  \bibfield  {author} {\bibinfo {author} {\bibfnamefont {Y.}~\bibnamefont
  {Li}}, \bibinfo {author} {\bibfnamefont {W.}~\bibnamefont {Hu}}, \ and\
  \bibinfo {author} {\bibfnamefont {M.}~\bibnamefont {Takada}},\ }\href
  {\doibase 10.1103/PhysRevD.89.083519} {\bibfield  {journal} {\bibinfo
  {journal} {Phys. Rev.}\ }\textbf {\bibinfo {volume} {D89}},\ \bibinfo {pages}
  {083519} (\bibinfo {year} {2014}{\natexlab{a}})},\ \Eprint
  {http://arxiv.org/abs/1401.0385} {arXiv:1401.0385 [astro-ph.CO]} \BibitemShut
  {NoStop}%
\bibitem [{\citenamefont {Chiang}\ \emph {et~al.}(2014)\citenamefont {Chiang},
  \citenamefont {Wagner}, \citenamefont {Schmidt},\ and\ \citenamefont
  {Komatsu}}]{Chiang:2014oga}%
  \BibitemOpen
  \bibfield  {author} {\bibinfo {author} {\bibfnamefont {C.-T.}\ \bibnamefont
  {Chiang}}, \bibinfo {author} {\bibfnamefont {C.}~\bibnamefont {Wagner}},
  \bibinfo {author} {\bibfnamefont {F.}~\bibnamefont {Schmidt}}, \ and\
  \bibinfo {author} {\bibfnamefont {E.}~\bibnamefont {Komatsu}},\ }\href
  {\doibase 10.1088/1475-7516/2014/05/048} {\bibfield  {journal} {\bibinfo
  {journal} {JCAP}\ }\textbf {\bibinfo {volume} {1405}},\ \bibinfo {pages}
  {048} (\bibinfo {year} {2014})},\ \Eprint {http://arxiv.org/abs/1403.3411}
  {arXiv:1403.3411 [astro-ph.CO]} \BibitemShut {NoStop}%
\bibitem [{\citenamefont {Manzotti}\ \emph {et~al.}(2014)\citenamefont
  {Manzotti}, \citenamefont {Hu},\ and\ \citenamefont
  {Benoit-Lévy}}]{Manzotti:2014wca}%
  \BibitemOpen
  \bibfield  {author} {\bibinfo {author} {\bibfnamefont {A.}~\bibnamefont
  {Manzotti}}, \bibinfo {author} {\bibfnamefont {W.}~\bibnamefont {Hu}}, \ and\
  \bibinfo {author} {\bibfnamefont {A.}~\bibnamefont {Benoit-Lévy}},\ }\href
  {\doibase 10.1103/PhysRevD.90.023003} {\bibfield  {journal} {\bibinfo
  {journal} {Phys. Rev.}\ }\textbf {\bibinfo {volume} {D90}},\ \bibinfo {pages}
  {023003} (\bibinfo {year} {2014})},\ \Eprint {http://arxiv.org/abs/1401.7992}
  {arXiv:1401.7992 [astro-ph.CO]} \BibitemShut {NoStop}%
\bibitem [{\citenamefont {Baldauf}\ \emph {et~al.}(2011)\citenamefont
  {Baldauf}, \citenamefont {Seljak}, \citenamefont {Senatore},\ and\
  \citenamefont {Zaldarriaga}}]{Baldauf:2011bh}%
  \BibitemOpen
  \bibfield  {author} {\bibinfo {author} {\bibfnamefont {T.}~\bibnamefont
  {Baldauf}}, \bibinfo {author} {\bibfnamefont {U.}~\bibnamefont {Seljak}},
  \bibinfo {author} {\bibfnamefont {L.}~\bibnamefont {Senatore}}, \ and\
  \bibinfo {author} {\bibfnamefont {M.}~\bibnamefont {Zaldarriaga}},\ }\href
  {\doibase 10.1088/1475-7516/2011/10/031} {\bibfield  {journal} {\bibinfo
  {journal} {JCAP}\ }\textbf {\bibinfo {volume} {1110}},\ \bibinfo {pages}
  {031} (\bibinfo {year} {2011})},\ \Eprint {http://arxiv.org/abs/1106.5507}
  {arXiv:1106.5507 [astro-ph.CO]} \BibitemShut {NoStop}%
\bibitem [{\citenamefont {Baldauf}\ \emph {et~al.}(2016)\citenamefont
  {Baldauf}, \citenamefont {Seljak}, \citenamefont {Senatore},\ and\
  \citenamefont {Zaldarriaga}}]{Baldauf:2015vio}%
  \BibitemOpen
  \bibfield  {author} {\bibinfo {author} {\bibfnamefont {T.}~\bibnamefont
  {Baldauf}}, \bibinfo {author} {\bibfnamefont {U.}~\bibnamefont {Seljak}},
  \bibinfo {author} {\bibfnamefont {L.}~\bibnamefont {Senatore}}, \ and\
  \bibinfo {author} {\bibfnamefont {M.}~\bibnamefont {Zaldarriaga}},\ }\href
  {\doibase 10.1088/1475-7516/2016/09/007} {\bibfield  {journal} {\bibinfo
  {journal} {JCAP}\ }\textbf {\bibinfo {volume} {1609}},\ \bibinfo {pages}
  {007} (\bibinfo {year} {2016})},\ \Eprint {http://arxiv.org/abs/1511.01465}
  {arXiv:1511.01465 [astro-ph.CO]} \BibitemShut {NoStop}%
\bibitem [{\citenamefont {Lazeyras}\ \emph {et~al.}(2016)\citenamefont
  {Lazeyras}, \citenamefont {Wagner}, \citenamefont {Baldauf},\ and\
  \citenamefont {Schmidt}}]{Lazeyras:2015lgp}%
  \BibitemOpen
  \bibfield  {author} {\bibinfo {author} {\bibfnamefont {T.}~\bibnamefont
  {Lazeyras}}, \bibinfo {author} {\bibfnamefont {C.}~\bibnamefont {Wagner}},
  \bibinfo {author} {\bibfnamefont {T.}~\bibnamefont {Baldauf}}, \ and\
  \bibinfo {author} {\bibfnamefont {F.}~\bibnamefont {Schmidt}},\ }\href
  {\doibase 10.1088/1475-7516/2016/02/018} {\bibfield  {journal} {\bibinfo
  {journal} {JCAP}\ }\textbf {\bibinfo {volume} {1602}},\ \bibinfo {pages}
  {018} (\bibinfo {year} {2016})},\ \Eprint {http://arxiv.org/abs/1511.01096}
  {arXiv:1511.01096 [astro-ph.CO]} \BibitemShut {NoStop}%
\bibitem [{\citenamefont {Gnedin}\ \emph {et~al.}(2011)\citenamefont {Gnedin},
  \citenamefont {Kravtsov},\ and\ \citenamefont {Rudd}}]{Gnedin:2011kj}%
  \BibitemOpen
  \bibfield  {author} {\bibinfo {author} {\bibfnamefont {N.~Y.}\ \bibnamefont
  {Gnedin}}, \bibinfo {author} {\bibfnamefont {A.~V.}\ \bibnamefont
  {Kravtsov}}, \ and\ \bibinfo {author} {\bibfnamefont {D.~H.}\ \bibnamefont
  {Rudd}},\ }\href {\doibase 10.1088/0067-0049/194/2/46} {\bibfield  {journal}
  {\bibinfo  {journal} {Astrophys. J. Suppl.}\ }\textbf {\bibinfo {volume}
  {194}},\ \bibinfo {pages} {46} (\bibinfo {year} {2011})},\ \Eprint
  {http://arxiv.org/abs/1104.1428} {arXiv:1104.1428 [astro-ph.CO]} \BibitemShut
  {NoStop}%
\bibitem [{\citenamefont {Li}\ \emph {et~al.}(2014{\natexlab{b}})\citenamefont
  {Li}, \citenamefont {Hu},\ and\ \citenamefont {Takada}}]{Li:2014jra}%
  \BibitemOpen
  \bibfield  {author} {\bibinfo {author} {\bibfnamefont {Y.}~\bibnamefont
  {Li}}, \bibinfo {author} {\bibfnamefont {W.}~\bibnamefont {Hu}}, \ and\
  \bibinfo {author} {\bibfnamefont {M.}~\bibnamefont {Takada}},\ }\href
  {\doibase 10.1103/PhysRevD.90.103530} {\bibfield  {journal} {\bibinfo
  {journal} {Phys. Rev.}\ }\textbf {\bibinfo {volume} {D90}},\ \bibinfo {pages}
  {103530} (\bibinfo {year} {2014}{\natexlab{b}})},\ \Eprint
  {http://arxiv.org/abs/1408.1081} {arXiv:1408.1081 [astro-ph.CO]} \BibitemShut
  {NoStop}%
\bibitem [{\citenamefont {Paranjape}\ and\ \citenamefont
  {Padmanabhan}(2017)}]{Paranjape:2016pbh}%
  \BibitemOpen
  \bibfield  {author} {\bibinfo {author} {\bibfnamefont {A.}~\bibnamefont
  {Paranjape}}\ and\ \bibinfo {author} {\bibfnamefont {N.}~\bibnamefont
  {Padmanabhan}},\ }\href {\doibase 10.1093/mnras/stx659} {\bibfield  {journal}
  {\bibinfo  {journal} {Mon. Not. Roy. Astron. Soc.}\ }\textbf {\bibinfo
  {volume} {468}},\ \bibinfo {pages} {2984} (\bibinfo {year} {2017})},\ \Eprint
  {http://arxiv.org/abs/1612.02833} {arXiv:1612.02833 [astro-ph.CO]}
  \BibitemShut {NoStop}%
\bibitem [{\citenamefont {Chiang}\ \emph {et~al.}(2016)\citenamefont {Chiang},
  \citenamefont {Li}, \citenamefont {Hu},\ and\ \citenamefont
  {LoVerde}}]{Chiang:2016vxa}%
  \BibitemOpen
  \bibfield  {author} {\bibinfo {author} {\bibfnamefont {C.-T.}\ \bibnamefont
  {Chiang}}, \bibinfo {author} {\bibfnamefont {Y.}~\bibnamefont {Li}}, \bibinfo
  {author} {\bibfnamefont {W.}~\bibnamefont {Hu}}, \ and\ \bibinfo {author}
  {\bibfnamefont {M.}~\bibnamefont {LoVerde}},\ }\href {\doibase
  10.1103/PhysRevD.94.123502} {\bibfield  {journal} {\bibinfo  {journal} {Phys.
  Rev.}\ }\textbf {\bibinfo {volume} {D94}},\ \bibinfo {pages} {123502}
  (\bibinfo {year} {2016})},\ \Eprint {http://arxiv.org/abs/1609.01701}
  {arXiv:1609.01701 [astro-ph.CO]} \BibitemShut {NoStop}%
\bibitem [{\citenamefont {Chiang}\ \emph {et~al.}(2017)\citenamefont {Chiang},
  \citenamefont {Hu}, \citenamefont {Li},\ and\ \citenamefont
  {Loverde}}]{Chiang:2017vuk}%
  \BibitemOpen
  \bibfield  {author} {\bibinfo {author} {\bibfnamefont {C.-T.}\ \bibnamefont
  {Chiang}}, \bibinfo {author} {\bibfnamefont {W.}~\bibnamefont {Hu}}, \bibinfo
  {author} {\bibfnamefont {Y.}~\bibnamefont {Li}}, \ and\ \bibinfo {author}
  {\bibfnamefont {M.}~\bibnamefont {Loverde}},\ }\href@noop {} {\  (\bibinfo
  {year} {2017})},\ \Eprint {http://arxiv.org/abs/1710.01310} {arXiv:1710.01310
  [astro-ph.CO]} \BibitemShut {NoStop}%
\bibitem [{\citenamefont {Armendariz-Picon}\ \emph {et~al.}(2000)\citenamefont
  {Armendariz-Picon}, \citenamefont {Mukhanov},\ and\ \citenamefont
  {Steinhardt}}]{ArmendarizPicon:2000dh}%
  \BibitemOpen
  \bibfield  {author} {\bibinfo {author} {\bibfnamefont {C.}~\bibnamefont
  {Armendariz-Picon}}, \bibinfo {author} {\bibfnamefont {V.~F.}\ \bibnamefont
  {Mukhanov}}, \ and\ \bibinfo {author} {\bibfnamefont {P.~J.}\ \bibnamefont
  {Steinhardt}},\ }\href {\doibase 10.1103/PhysRevLett.85.4438} {\bibfield
  {journal} {\bibinfo  {journal} {Phys. Rev. Lett.}\ }\textbf {\bibinfo
  {volume} {85}},\ \bibinfo {pages} {4438} (\bibinfo {year} {2000})},\ \Eprint
  {http://arxiv.org/abs/astro-ph/0004134} {arXiv:astro-ph/0004134 [astro-ph]}
  \BibitemShut {NoStop}%
\bibitem [{\citenamefont {Garriga}\ and\ \citenamefont
  {Mukhanov}(1999)}]{Garriga:1999vw}%
  \BibitemOpen
  \bibfield  {author} {\bibinfo {author} {\bibfnamefont {J.}~\bibnamefont
  {Garriga}}\ and\ \bibinfo {author} {\bibfnamefont {V.~F.}\ \bibnamefont
  {Mukhanov}},\ }\href {\doibase 10.1016/S0370-2693(99)00602-4} {\bibfield
  {journal} {\bibinfo  {journal} {Phys. Lett.}\ }\textbf {\bibinfo {volume}
  {B458}},\ \bibinfo {pages} {219} (\bibinfo {year} {1999})},\ \Eprint
  {http://arxiv.org/abs/hep-th/9904176} {arXiv:hep-th/9904176 [hep-th]}
  \BibitemShut {NoStop}%
\bibitem [{\citenamefont {Gordon}\ and\ \citenamefont
  {Hu}(2004)}]{Gordon:2004ez}%
  \BibitemOpen
  \bibfield  {author} {\bibinfo {author} {\bibfnamefont {C.}~\bibnamefont
  {Gordon}}\ and\ \bibinfo {author} {\bibfnamefont {W.}~\bibnamefont {Hu}},\
  }\href {\doibase 10.1103/PhysRevD.70.083003} {\bibfield  {journal} {\bibinfo
  {journal} {Phys. Rev.}\ }\textbf {\bibinfo {volume} {D70}},\ \bibinfo {pages}
  {083003} (\bibinfo {year} {2004})},\ \Eprint
  {http://arxiv.org/abs/astro-ph/0406496} {arXiv:astro-ph/0406496 [astro-ph]}
  \BibitemShut {NoStop}%
\bibitem [{\citenamefont {Bardeen}(1980)}]{PhysRevD.22.1882}%
  \BibitemOpen
  \bibfield  {author} {\bibinfo {author} {\bibfnamefont {J.~M.}\ \bibnamefont
  {Bardeen}},\ }\href {\doibase 10.1103/PhysRevD.22.1882} {\bibfield  {journal}
  {\bibinfo  {journal} {Phys. Rev. D}\ }\textbf {\bibinfo {volume} {22}},\
  \bibinfo {pages} {1882} (\bibinfo {year} {1980})}\BibitemShut {NoStop}%
\bibitem [{\citenamefont {Kodama}\ and\ \citenamefont
  {Sasaki}(1984)}]{Kodama:1984PTPS}%
  \BibitemOpen
  \bibfield  {author} {\bibinfo {author} {\bibfnamefont {H.}~\bibnamefont
  {Kodama}}\ and\ \bibinfo {author} {\bibfnamefont {M.}~\bibnamefont
  {Sasaki}},\ }\href {\doibase 10.1143/PTPS.78.1} {\bibfield  {journal}
  {\bibinfo  {journal} {Progress of Theoretical Physics Supplement}\ }\textbf
  {\bibinfo {volume} {78}},\ \bibinfo {pages} {1} (\bibinfo {year}
  {1984})}\BibitemShut {NoStop}%
\bibitem [{\citenamefont {Takada}\ and\ \citenamefont
  {Hu}(2013)}]{Takada:2013bfn}%
  \BibitemOpen
  \bibfield  {author} {\bibinfo {author} {\bibfnamefont {M.}~\bibnamefont
  {Takada}}\ and\ \bibinfo {author} {\bibfnamefont {W.}~\bibnamefont {Hu}},\
  }\href {\doibase 10.1103/PhysRevD.87.123504} {\bibfield  {journal} {\bibinfo
  {journal} {Phys. Rev.}\ }\textbf {\bibinfo {volume} {D87}},\ \bibinfo {pages}
  {123504} (\bibinfo {year} {2013})},\ \Eprint {http://arxiv.org/abs/1302.6994}
  {arXiv:1302.6994 [astro-ph.CO]} \BibitemShut {NoStop}%
\bibitem [{\citenamefont {Ma}(2007)}]{Ma:2006zk}%
  \BibitemOpen
  \bibfield  {author} {\bibinfo {author} {\bibfnamefont {Z.-M.}\ \bibnamefont
  {Ma}},\ }\href {\doibase 10.1086/519440} {\bibfield  {journal} {\bibinfo
  {journal} {Astrophys. J.}\ }\textbf {\bibinfo {volume} {665}},\ \bibinfo
  {pages} {887} (\bibinfo {year} {2007})},\ \Eprint
  {http://arxiv.org/abs/astro-ph/0610213} {arXiv:astro-ph/0610213 [astro-ph]}
  \BibitemShut {NoStop}%
\bibitem [{\citenamefont {Valageas}(2014)}]{Valageas:2013zda}%
  \BibitemOpen
  \bibfield  {author} {\bibinfo {author} {\bibfnamefont {P.}~\bibnamefont
  {Valageas}},\ }\href {\doibase 10.1103/PhysRevD.89.123522} {\bibfield
  {journal} {\bibinfo  {journal} {Phys. Rev.}\ }\textbf {\bibinfo {volume}
  {D89}},\ \bibinfo {pages} {123522} (\bibinfo {year} {2014})},\ \Eprint
  {http://arxiv.org/abs/1311.4286} {arXiv:1311.4286 [astro-ph.CO]} \BibitemShut
  {NoStop}%
\bibitem [{\citenamefont {Springel}(2005)}]{Springel:2005mi}%
  \BibitemOpen
  \bibfield  {author} {\bibinfo {author} {\bibfnamefont {V.}~\bibnamefont
  {Springel}},\ }\href {\doibase 10.1111/j.1365-2966.2005.09655.x} {\bibfield
  {journal} {\bibinfo  {journal} {Mon. Not. Roy. Astron. Soc.}\ }\textbf
  {\bibinfo {volume} {364}},\ \bibinfo {pages} {1105} (\bibinfo {year}
  {2005})},\ \Eprint {http://arxiv.org/abs/astro-ph/0505010}
  {arXiv:astro-ph/0505010 [astro-ph]} \BibitemShut {NoStop}%
\bibitem [{\citenamefont {Blas}\ \emph {et~al.}(2011)\citenamefont {Blas},
  \citenamefont {Lesgourgues},\ and\ \citenamefont {Tram}}]{Blas:2011rf}%
  \BibitemOpen
  \bibfield  {author} {\bibinfo {author} {\bibfnamefont {D.}~\bibnamefont
  {Blas}}, \bibinfo {author} {\bibfnamefont {J.}~\bibnamefont {Lesgourgues}}, \
  and\ \bibinfo {author} {\bibfnamefont {T.}~\bibnamefont {Tram}},\ }\href
  {\doibase 10.1088/1475-7516/2011/07/034} {\bibfield  {journal} {\bibinfo
  {journal} {JCAP}\ }\textbf {\bibinfo {volume} {1107}},\ \bibinfo {pages}
  {034} (\bibinfo {year} {2011})},\ \Eprint {http://arxiv.org/abs/1104.2933}
  {arXiv:1104.2933 [astro-ph.CO]} \BibitemShut {NoStop}%
\bibitem [{\citenamefont {Crocce}\ \emph {et~al.}(2006)\citenamefont {Crocce},
  \citenamefont {Pueblas},\ and\ \citenamefont {Scoccimarro}}]{Crocce:2006ve}%
  \BibitemOpen
  \bibfield  {author} {\bibinfo {author} {\bibfnamefont {M.}~\bibnamefont
  {Crocce}}, \bibinfo {author} {\bibfnamefont {S.}~\bibnamefont {Pueblas}}, \
  and\ \bibinfo {author} {\bibfnamefont {R.}~\bibnamefont {Scoccimarro}},\
  }\href {\doibase 10.1111/j.1365-2966.2006.11040.x} {\bibfield  {journal}
  {\bibinfo  {journal} {Mon. Not. Roy. Astron. Soc.}\ }\textbf {\bibinfo
  {volume} {373}},\ \bibinfo {pages} {369} (\bibinfo {year} {2006})},\ \Eprint
  {http://arxiv.org/abs/astro-ph/0606505} {arXiv:astro-ph/0606505 [astro-ph]}
  \BibitemShut {NoStop}%
\bibitem [{\citenamefont {Frigo}\ and\ \citenamefont {Johnson}(2005)}]{FFTW05}%
  \BibitemOpen
  \bibfield  {author} {\bibinfo {author} {\bibfnamefont {M.}~\bibnamefont
  {Frigo}}\ and\ \bibinfo {author} {\bibfnamefont {S.~G.}\ \bibnamefont
  {Johnson}},\ }\href@noop {} {\bibfield  {journal} {\bibinfo  {journal}
  {Proceedings of the IEEE}\ }\textbf {\bibinfo {volume} {93}},\ \bibinfo
  {pages} {216} (\bibinfo {year} {2005})},\ \bibinfo {note} {special issue on
  ``Program Generation, Optimization, and Platform Adaptation''}\BibitemShut
  {NoStop}%
\bibitem [{\citenamefont {Jeong}\ and\ \citenamefont
  {Komatsu}(2006)}]{Jeong:2006xd}%
  \BibitemOpen
  \bibfield  {author} {\bibinfo {author} {\bibfnamefont {D.}~\bibnamefont
  {Jeong}}\ and\ \bibinfo {author} {\bibfnamefont {E.}~\bibnamefont
  {Komatsu}},\ }\href {\doibase 10.1086/507781} {\bibfield  {journal} {\bibinfo
   {journal} {Astrophys. J.}\ }\textbf {\bibinfo {volume} {651}},\ \bibinfo
  {pages} {619} (\bibinfo {year} {2006})},\ \Eprint
  {http://arxiv.org/abs/astro-ph/0604075} {arXiv:astro-ph/0604075 [astro-ph]}
  \BibitemShut {NoStop}%
\bibitem [{\citenamefont {Li}\ \emph {et~al.}(2016)\citenamefont {Li},
  \citenamefont {Hu},\ and\ \citenamefont {Takada}}]{Li:2015jsz}%
  \BibitemOpen
  \bibfield  {author} {\bibinfo {author} {\bibfnamefont {Y.}~\bibnamefont
  {Li}}, \bibinfo {author} {\bibfnamefont {W.}~\bibnamefont {Hu}}, \ and\
  \bibinfo {author} {\bibfnamefont {M.}~\bibnamefont {Takada}},\ }\href
  {\doibase 10.1103/PhysRevD.93.063507} {\bibfield  {journal} {\bibinfo
  {journal} {Phys. Rev.}\ }\textbf {\bibinfo {volume} {D93}},\ \bibinfo {pages}
  {063507} (\bibinfo {year} {2016})},\ \Eprint
  {http://arxiv.org/abs/1511.01454} {arXiv:1511.01454 [astro-ph.CO]}
  \BibitemShut {NoStop}%
\bibitem [{\citenamefont {Behroozi}\ \emph {et~al.}(2013)\citenamefont
  {Behroozi}, \citenamefont {Wechsler},\ and\ \citenamefont
  {Wu}}]{Behroozi:2011ju}%
  \BibitemOpen
  \bibfield  {author} {\bibinfo {author} {\bibfnamefont {P.~S.}\ \bibnamefont
  {Behroozi}}, \bibinfo {author} {\bibfnamefont {R.~H.}\ \bibnamefont
  {Wechsler}}, \ and\ \bibinfo {author} {\bibfnamefont {H.-Y.}\ \bibnamefont
  {Wu}},\ }\href {\doibase 10.1088/0004-637X/762/2/109} {\bibfield  {journal}
  {\bibinfo  {journal} {Astrophys. J.}\ }\textbf {\bibinfo {volume} {762}},\
  \bibinfo {pages} {109} (\bibinfo {year} {2013})},\ \Eprint
  {http://arxiv.org/abs/1110.4372} {arXiv:1110.4372 [astro-ph.CO]} \BibitemShut
  {NoStop}%
\bibitem [{\citenamefont {Dvali}\ \emph
  {et~al.}(2004{\natexlab{a}})\citenamefont {Dvali}, \citenamefont {Gruzinov},\
  and\ \citenamefont {Zaldarriaga}}]{Dvali:2003ar}%
  \BibitemOpen
  \bibfield  {author} {\bibinfo {author} {\bibfnamefont {G.}~\bibnamefont
  {Dvali}}, \bibinfo {author} {\bibfnamefont {A.}~\bibnamefont {Gruzinov}}, \
  and\ \bibinfo {author} {\bibfnamefont {M.}~\bibnamefont {Zaldarriaga}},\
  }\href {\doibase 10.1103/PhysRevD.69.083505} {\bibfield  {journal} {\bibinfo
  {journal} {Phys. Rev.}\ }\textbf {\bibinfo {volume} {D69}},\ \bibinfo {pages}
  {083505} (\bibinfo {year} {2004}{\natexlab{a}})},\ \Eprint
  {http://arxiv.org/abs/astro-ph/0305548} {arXiv:astro-ph/0305548 [astro-ph]}
  \BibitemShut {NoStop}%
\bibitem [{\citenamefont {{Desjacques}}\ \emph {et~al.}(2010)\citenamefont
  {{Desjacques}}, \citenamefont {{Crocce}}, \citenamefont {{Scoccimarro}},\
  and\ \citenamefont {{Sheth}}}]{2010PhRvD..82j3529D}%
  \BibitemOpen
  \bibfield  {author} {\bibinfo {author} {\bibfnamefont {V.}~\bibnamefont
  {{Desjacques}}}, \bibinfo {author} {\bibfnamefont {M.}~\bibnamefont
  {{Crocce}}}, \bibinfo {author} {\bibfnamefont {R.}~\bibnamefont
  {{Scoccimarro}}}, \ and\ \bibinfo {author} {\bibfnamefont {R.~K.}\
  \bibnamefont {{Sheth}}},\ }\href {\doibase 10.1103/PhysRevD.82.103529}
  {\bibfield  {journal} {\bibinfo  {journal} {\prd}\ }\textbf {\bibinfo
  {volume} {82}},\ \bibinfo {eid} {103529} (\bibinfo {year} {2010})},\ \Eprint
  {http://arxiv.org/abs/1009.3449} {arXiv:1009.3449 [astro-ph.CO]} \BibitemShut
  {NoStop}%
\bibitem [{\citenamefont {{Assassi}}\ \emph {et~al.}(2014)\citenamefont
  {{Assassi}}, \citenamefont {{Baumann}}, \citenamefont {{Green}},\ and\
  \citenamefont {{Zaldarriaga}}}]{2014JCAP...08..056A}%
  \BibitemOpen
  \bibfield  {author} {\bibinfo {author} {\bibfnamefont {V.}~\bibnamefont
  {{Assassi}}}, \bibinfo {author} {\bibfnamefont {D.}~\bibnamefont
  {{Baumann}}}, \bibinfo {author} {\bibfnamefont {D.}~\bibnamefont {{Green}}},
  \ and\ \bibinfo {author} {\bibfnamefont {M.}~\bibnamefont {{Zaldarriaga}}},\
  }\href {\doibase 10.1088/1475-7516/2014/08/056} {\bibfield  {journal}
  {\bibinfo  {journal} {Journal of Cosmology and Astro-Particle Physics}\
  }\textbf {\bibinfo {volume} {2014}},\ \bibinfo {eid} {056} (\bibinfo {year}
  {2014})},\ \Eprint {http://arxiv.org/abs/1402.5916} {arXiv:1402.5916
  [astro-ph.CO]} \BibitemShut {NoStop}%
\bibitem [{\citenamefont {Hui}\ and\ \citenamefont
  {Parfrey}(2008)}]{Hui:2007zh}%
  \BibitemOpen
  \bibfield  {author} {\bibinfo {author} {\bibfnamefont {L.}~\bibnamefont
  {Hui}}\ and\ \bibinfo {author} {\bibfnamefont {K.~P.}\ \bibnamefont
  {Parfrey}},\ }\href {\doibase 10.1103/PhysRevD.77.043527} {\bibfield
  {journal} {\bibinfo  {journal} {Phys. Rev.}\ }\textbf {\bibinfo {volume}
  {D77}},\ \bibinfo {pages} {043527} (\bibinfo {year} {2008})},\ \Eprint
  {http://arxiv.org/abs/0712.1162} {arXiv:0712.1162 [astro-ph]} \BibitemShut
  {NoStop}%
\bibitem [{\citenamefont {Parfrey}\ \emph {et~al.}(2011)\citenamefont
  {Parfrey}, \citenamefont {Hui},\ and\ \citenamefont
  {Sheth}}]{Parfrey:2010uy}%
  \BibitemOpen
  \bibfield  {author} {\bibinfo {author} {\bibfnamefont {K.}~\bibnamefont
  {Parfrey}}, \bibinfo {author} {\bibfnamefont {L.}~\bibnamefont {Hui}}, \ and\
  \bibinfo {author} {\bibfnamefont {R.~K.}\ \bibnamefont {Sheth}},\ }\href
  {\doibase 10.1103/PhysRevD.83.063511} {\bibfield  {journal} {\bibinfo
  {journal} {Phys. Rev.}\ }\textbf {\bibinfo {volume} {D83}},\ \bibinfo {pages}
  {063511} (\bibinfo {year} {2011})},\ \Eprint {http://arxiv.org/abs/1012.1335}
  {arXiv:1012.1335 [astro-ph.CO]} \BibitemShut {NoStop}%
\bibitem [{\citenamefont {LoVerde}(2014)}]{LoVerde:2014pxa}%
  \BibitemOpen
  \bibfield  {author} {\bibinfo {author} {\bibfnamefont {M.}~\bibnamefont
  {LoVerde}},\ }\href {\doibase 10.1103/PhysRevD.90.083530} {\bibfield
  {journal} {\bibinfo  {journal} {Phys. Rev.}\ }\textbf {\bibinfo {volume}
  {D90}},\ \bibinfo {pages} {083530} (\bibinfo {year} {2014})},\ \Eprint
  {http://arxiv.org/abs/1405.4855} {arXiv:1405.4855 [astro-ph.CO]} \BibitemShut
  {NoStop}%
\bibitem [{\citenamefont {Chiang}\ \emph {et~al.}(2018)\citenamefont {Chiang},
  \citenamefont {LoVerde},\ and\ \citenamefont
  {Villaescusa-Navarro}}]{Chiang:2018laa}%
  \BibitemOpen
  \bibfield  {author} {\bibinfo {author} {\bibfnamefont {C.-T.}\ \bibnamefont
  {Chiang}}, \bibinfo {author} {\bibfnamefont {M.}~\bibnamefont {LoVerde}}, \
  and\ \bibinfo {author} {\bibfnamefont {F.}~\bibnamefont
  {Villaescusa-Navarro}},\ }\href@noop {} {\  (\bibinfo {year} {2018})},\
  \Eprint {http://arxiv.org/abs/1811.12412} {arXiv:1811.12412 [astro-ph.CO]}
  \BibitemShut {NoStop}%
\bibitem [{\citenamefont {Matsubara}(1999)}]{Matsubara:1999qq}%
  \BibitemOpen
  \bibfield  {author} {\bibinfo {author} {\bibfnamefont {T.}~\bibnamefont
  {Matsubara}},\ }\href {\doibase 10.1086/307931} {\bibfield  {journal}
  {\bibinfo  {journal} {Astrophys. J.}\ }\textbf {\bibinfo {volume} {525}},\
  \bibinfo {pages} {543} (\bibinfo {year} {1999})},\ \Eprint
  {http://arxiv.org/abs/astro-ph/9906029} {arXiv:astro-ph/9906029 [astro-ph]}
  \BibitemShut {NoStop}%
\bibitem [{\citenamefont {Smith}\ and\ \citenamefont
  {LoVerde}(2011)}]{Smith:2010gx}%
  \BibitemOpen
  \bibfield  {author} {\bibinfo {author} {\bibfnamefont {K.~M.}\ \bibnamefont
  {Smith}}\ and\ \bibinfo {author} {\bibfnamefont {M.}~\bibnamefont
  {LoVerde}},\ }\href {\doibase 10.1088/1475-7516/2011/11/009} {\bibfield
  {journal} {\bibinfo  {journal} {JCAP}\ }\textbf {\bibinfo {volume} {1111}},\
  \bibinfo {pages} {009} (\bibinfo {year} {2011})},\ \Eprint
  {http://arxiv.org/abs/1010.0055} {arXiv:1010.0055 [astro-ph.CO]} \BibitemShut
  {NoStop}%
\bibitem [{\citenamefont {Tegmark}\ and\ \citenamefont
  {Zaldarriaga}(2002)}]{Tegmark:2002cy}%
  \BibitemOpen
  \bibfield  {author} {\bibinfo {author} {\bibfnamefont {M.}~\bibnamefont
  {Tegmark}}\ and\ \bibinfo {author} {\bibfnamefont {M.}~\bibnamefont
  {Zaldarriaga}},\ }\href {\doibase 10.1103/PhysRevD.66.103508} {\bibfield
  {journal} {\bibinfo  {journal} {Phys. Rev.}\ }\textbf {\bibinfo {volume}
  {D66}},\ \bibinfo {pages} {103508} (\bibinfo {year} {2002})},\ \Eprint
  {http://arxiv.org/abs/astro-ph/0207047} {arXiv:astro-ph/0207047 [astro-ph]}
  \BibitemShut {NoStop}%
\bibitem [{\citenamefont {{Hlozek}}\ \emph {et~al.}(2012)\citenamefont
  {{Hlozek}} \emph {et~al.}}]{2012ApJ...749...90H}%
  \BibitemOpen
  \bibfield  {author} {\bibinfo {author} {\bibfnamefont {R.}~\bibnamefont
  {{Hlozek}}} \emph {et~al.},\ }\href {\doibase 10.1088/0004-637X/749/1/90}
  {\bibfield  {journal} {\bibinfo  {journal} {\apj}\ }\textbf {\bibinfo
  {volume} {749}},\ \bibinfo {eid} {90} (\bibinfo {year} {2012})},\ \Eprint
  {http://arxiv.org/abs/1105.4887} {arXiv:1105.4887} \BibitemShut {NoStop}%
\bibitem [{\citenamefont {Kaiser}(1987)}]{Kaiser:1987qv}%
  \BibitemOpen
  \bibfield  {author} {\bibinfo {author} {\bibfnamefont {N.}~\bibnamefont
  {Kaiser}},\ }\href@noop {} {\bibfield  {journal} {\bibinfo  {journal} {Mon.
  Not. Roy. Astron. Soc.}\ }\textbf {\bibinfo {volume} {227}},\ \bibinfo
  {pages} {1} (\bibinfo {year} {1987})}\BibitemShut {NoStop}%
\bibitem [{\citenamefont {Liu}\ \emph {et~al.}(2011)\citenamefont {Liu},
  \citenamefont {Li},\ and\ \citenamefont {Zhang}}]{Liu:2010ba}%
  \BibitemOpen
  \bibfield  {author} {\bibinfo {author} {\bibfnamefont {J.}~\bibnamefont
  {Liu}}, \bibinfo {author} {\bibfnamefont {M.}~\bibnamefont {Li}}, \ and\
  \bibinfo {author} {\bibfnamefont {X.}~\bibnamefont {Zhang}},\ }\href
  {\doibase 10.1088/1475-7516/2011/06/028} {\bibfield  {journal} {\bibinfo
  {journal} {JCAP}\ }\textbf {\bibinfo {volume} {1106}},\ \bibinfo {pages}
  {028} (\bibinfo {year} {2011})},\ \Eprint {http://arxiv.org/abs/1011.6146}
  {arXiv:1011.6146 [astro-ph.CO]} \BibitemShut {NoStop}%
\bibitem [{\citenamefont {Dvali}\ \emph
  {et~al.}(2004{\natexlab{b}})\citenamefont {Dvali}, \citenamefont {Gruzinov},\
  and\ \citenamefont {Zaldarriaga}}]{Dvali:2003em}%
  \BibitemOpen
  \bibfield  {author} {\bibinfo {author} {\bibfnamefont {G.}~\bibnamefont
  {Dvali}}, \bibinfo {author} {\bibfnamefont {A.}~\bibnamefont {Gruzinov}}, \
  and\ \bibinfo {author} {\bibfnamefont {M.}~\bibnamefont {Zaldarriaga}},\
  }\href {\doibase 10.1103/PhysRevD.69.023505} {\bibfield  {journal} {\bibinfo
  {journal} {Phys. Rev.}\ }\textbf {\bibinfo {volume} {D69}},\ \bibinfo {pages}
  {023505} (\bibinfo {year} {2004}{\natexlab{b}})},\ \Eprint
  {http://arxiv.org/abs/astro-ph/0303591} {arXiv:astro-ph/0303591 [astro-ph]}
  \BibitemShut {NoStop}%
\bibitem [{\citenamefont {Linde}\ and\ \citenamefont
  {Mukhanov}(1997)}]{Linde:1996gt}%
  \BibitemOpen
  \bibfield  {author} {\bibinfo {author} {\bibfnamefont {A.~D.}\ \bibnamefont
  {Linde}}\ and\ \bibinfo {author} {\bibfnamefont {V.~F.}\ \bibnamefont
  {Mukhanov}},\ }\href {\doibase 10.1103/PhysRevD.56.R535} {\bibfield
  {journal} {\bibinfo  {journal} {Phys. Rev.}\ }\textbf {\bibinfo {volume}
  {D56}},\ \bibinfo {pages} {R535} (\bibinfo {year} {1997})},\ \Eprint
  {http://arxiv.org/abs/astro-ph/9610219} {arXiv:astro-ph/9610219 [astro-ph]}
  \BibitemShut {NoStop}%
\bibitem [{\citenamefont {Lyth}\ and\ \citenamefont
  {Wands}(2002)}]{Lyth:2001nq}%
  \BibitemOpen
  \bibfield  {author} {\bibinfo {author} {\bibfnamefont {D.~H.}\ \bibnamefont
  {Lyth}}\ and\ \bibinfo {author} {\bibfnamefont {D.}~\bibnamefont {Wands}},\
  }\href {\doibase 10.1016/S0370-2693(01)01366-1} {\bibfield  {journal}
  {\bibinfo  {journal} {Phys. Lett.}\ }\textbf {\bibinfo {volume} {B524}},\
  \bibinfo {pages} {5} (\bibinfo {year} {2002})},\ \Eprint
  {http://arxiv.org/abs/hep-ph/0110002} {arXiv:hep-ph/0110002 [hep-ph]}
  \BibitemShut {NoStop}%
\bibitem [{\citenamefont {Moroi}\ and\ \citenamefont
  {Takahashi}(2001)}]{Moroi:2001ct}%
  \BibitemOpen
  \bibfield  {author} {\bibinfo {author} {\bibfnamefont {T.}~\bibnamefont
  {Moroi}}\ and\ \bibinfo {author} {\bibfnamefont {T.}~\bibnamefont
  {Takahashi}},\ }\href {\doibase 10.1016/S0370-2693(02)02070-1,
  10.1016/S0370-2693(01)01295-3} {\bibfield  {journal} {\bibinfo  {journal}
  {Phys. Lett.}\ }\textbf {\bibinfo {volume} {B522}},\ \bibinfo {pages} {215}
  (\bibinfo {year} {2001})},\ \bibinfo {note} {[Erratum: Phys.
  Lett.B539,303(2002)]},\ \Eprint {http://arxiv.org/abs/hep-ph/0110096}
  {arXiv:hep-ph/0110096 [hep-ph]} \BibitemShut {NoStop}%
\bibitem [{\citenamefont {Enqvist}\ and\ \citenamefont
  {Sloth}(2002)}]{Enqvist:2001zp}%
  \BibitemOpen
  \bibfield  {author} {\bibinfo {author} {\bibfnamefont {K.}~\bibnamefont
  {Enqvist}}\ and\ \bibinfo {author} {\bibfnamefont {M.~S.}\ \bibnamefont
  {Sloth}},\ }\href {\doibase 10.1016/S0550-3213(02)00043-3} {\bibfield
  {journal} {\bibinfo  {journal} {Nucl. Phys.}\ }\textbf {\bibinfo {volume}
  {B626}},\ \bibinfo {pages} {395} (\bibinfo {year} {2002})},\ \Eprint
  {http://arxiv.org/abs/hep-ph/0109214} {arXiv:hep-ph/0109214 [hep-ph]}
  \BibitemShut {NoStop}%
\bibitem [{\citenamefont {Fujita}\ \emph {et~al.}(2015)\citenamefont {Fujita},
  \citenamefont {Yokoyama},\ and\ \citenamefont {Yokoyama}}]{Fujita:2014oba}%
  \BibitemOpen
  \bibfield  {author} {\bibinfo {author} {\bibfnamefont {T.}~\bibnamefont
  {Fujita}}, \bibinfo {author} {\bibfnamefont {J.}~\bibnamefont {Yokoyama}}, \
  and\ \bibinfo {author} {\bibfnamefont {S.}~\bibnamefont {Yokoyama}},\ }\href
  {\doibase 10.1093/ptep/ptv037} {\bibfield  {journal} {\bibinfo  {journal}
  {PTEP}\ }\textbf {\bibinfo {volume} {2015}},\ \bibinfo {pages} {043E01}
  (\bibinfo {year} {2015})},\ \Eprint {http://arxiv.org/abs/1411.3658}
  {arXiv:1411.3658 [astro-ph.CO]} \BibitemShut {NoStop}%
\end{thebibliography}%
	
\appendix*
\section{On the generation of quintessence isocurvature perturbations}
\label{sec:generationmodes}

	In this appendix we will make some basic observations about the possibility of generating quintessence isocurvature perturbations that are large enough, in comparison with adiabatic perturbations, to have observational impact at late times. 
	
	Dark energy isocurvature perturbations that are perfectly anticorrelated with the primordial curvature perturbations have been studied by \cite{Gordon:2004ez}, among others, for the purposes of producing a low quadrupole amplitude in the CMB temperature power spectrum. The mechanism studied involves a delayed, postinflationary production of curvature perturbations sourced by the same field that sources the dark energy isocurvature perturbations. This is achieved, for instance, by making the quintessence field the field that modulates reheating \cite{Dvali:2003em,Dvali:2003ar} or via curvatons \cite{Linde:1996gt, Lyth:2001nq, Moroi:2001ct, Enqvist:2001zp}.
	
	 In \cite{Gordon:2004ez}, the authors argue that inflationary generation of anticorrelated quintessence isocurvature perturbations generally makes the tensor-to-scalar ratio too large. However, in \cite{Gordon:2004ez} the scalar perturbations sourced by the inflaton were neglected. If included, they keep the tensor-to-scalar ratio small but will unavoidably dominate over the scalar curvature perturbations generated after inflation. This necessarily makes the isocurvature perturbations small and only weakly correlated to the adiabatic modes.  On the other hand, we find it is generally possible to satisfy these criteria with a modified kinetic term, including the one used in the main text of this paper [see Eq.~\ref{eq:L})]. 
	  
	Consider a scalar field $Q$ with a Lagrangian that is a general function of $X$ and $Q$, 
	\begin{align}
		\mathcal{L} = P(X,Q) \, , 
	\end{align}
	where
	\begin{align}
		X = -\frac{1}{2}g^{\mu\nu}\partial_\mu Q\partial_\nu Q \, . 
	\end{align}
	The energy density and pressure are
	\begin{align}
		\rho_Q & = 2XP_{,X} -P\,, \\
		p_Q & =P \, .
	\end{align}
	For a homogeneous background, we have
	\begin{align}
		\dot{\rho}_Q= -3H(\rho_Q+p_Q)\,,
	\end{align}
	which gives equation of motion for $Q$
	\begin{align}
		\ddot{Q}+ 3c_Q^2 H\dot{Q}+c^2_Q\dot{Q}^2\frac{P_{,XQ}}{P_{,X}}-c_Q^2\frac{P_{,Q}}{P_{,X}} =0\,,
	\end{align}
	where the effective sound speed is given by
	\begin{align}
		\label{eq:cQ}
		c_Q^2 = \frac{{P}_{,X}}{{P}_{,X} + 2{P}_{,XX}{X}} \, .
	\end{align}
	It will also be convenient to define
	\begin{align}
		\label{eq:cX}
		c_X^2 =\frac{{P}_{,Q}}{2X P_{,XQ}-P_{,Q}}\, .
	\end{align}
	In absence of derivative couplings of the field to itself, $P_{,XQ} = 0$, so we have $c_X^2=-1$.
	
	\subsection{$Q$ as dark energy}
	\label{ssec:QasDE}
	
	We define slow-roll parameters for $Q$ as
	\begin{align}
		\label{eq:epsilonQ}
		\epsilon_Q & \equiv -\frac{1}{2H}\frac{\dot\rho_Q}{\rho_Q}\,,\\
		\eta_Q & \equiv \frac{\dot\epsilon_Q}{H\epsilon_Q}\, ,
	\end{align}
	which can be rewritten as
	\begin{align}
		\label{eq:epsilonQofP}
		\epsilon_Q &= \frac{3XP_{,X}}{2XP_{,X}-P} \, ,\\
		\label{eq:etaQ}	
		\eta_Q &= \frac{\dot{X}}{HX}\left(\frac{c_Q^2 +1}{2c_Q^2}\right) + 2\epsilon_Q+ \frac{\dot{Q}P_{,QX}}{HP_{,X}} \, .
	\end{align}
	In the limit $\epsilon_Q\ll 1$, we have $w_Q\approx -1$. The additional condition $\eta_Q \ll 1$ can be imposed to keep $\epsilon_Q\ll1$ but is not necessary for our purposes. In particular, the example solution used in the main text of the paper has  $\eta_Q\approx \frac{3}{2}(1+c_Q^2)(1+w_{tot})\sim \mathcal{O}(1)$ during the matter and radiation eras while $\epsilon_Q\ll1$ throughout. 

	In what follows it will be convenient to rewrite the continuity equation in terms of the sounds speeds in Eq.~(\ref{eq:cQ}) and Eq.~(\ref{eq:cX}) and the slow-roll parameters in Eq.~(\ref{eq:epsilonQofP}) and Eq.~(\ref{eq:etaQ}). This gives an expression for the change in energy due to change in the field strength as
	\begin{align}
		\label{eq:drho}
		\frac{\partial\log\rho_Q}{\partial Q} = -\frac{2\epsilon_Q}{Q'}\lb\frac{c_Q^2+1}{c_Q^2 - c_X^2}\rb\lb1 + \frac{\eta_Q -2 \epsilon_Q}{3(1+c_Q^2)}\rb \, .
	\end{align}

	\subsection{Inflationary generation of $\delta Q$}	
	\label{ssec:deltaQ} 

	We assume that the field $Q$ is a spectator field during inflation and that mixing between the inflaton and $Q$ can be ignored. The field $Q$ then acquires perturbations $\delta Q$ that will have power spectrum \cite{Fujita:2014oba},
	\begin{align}
		\Delta_{Q}^2  = \frac{H_{i}^2}{c_{Qi} \bar{P}_{,Xi}4\pi^2} \, ,
	\end{align}
	where the subscript $i$ indicates these quantities are to be evaluated during inflation. The $\delta Q$ will be our isocurvature modes. For the isocurvature modes to be correlated with $\R$, we assume at some time after inflation fluctuations in the energy density of $Q$ create adiabatic fluctuations via some mechanism with an efficiency factor $\gamma$, 
	\begin{align}
		\label{eq:RQ}
		\R_{Q} \equiv \gamma \frac{\delta Q}{Q_i} \, ,
	\end{align}
	where $Q_i$ is the value of the field at the end of inflation. The subscript $Q$ in Eq.~(\ref{eq:RQ}) indicates that these perturbations were inherited from $\delta Q$ after inflation. 
	
	In addition, there will be adiabatic perturbations generated during inflation, we label these perturbations $\R_{i}$. For simplicity we assume slow-roll inflation so that 
	\begin{align}
		\R_{i} = -\frac{\delta \phi}{\phi'} \, ,
	\end{align}
	where $\phi$ is the inflaton and the power spectrum of $\delta \phi$ is 
	\begin{align}
		\Delta^2_{\phi} =\frac{H_{i}^2}{4\pi^2} \, ,
	\end{align}
	so that
	\begin{align}
		\Delta^2_{\R_{i}} =\frac{H_{i}^2}{8\pi^2M_{pl}^2\epsilon_{i}} \, ,
	\end{align}
	where $\epsilon_{i}$ is the usual inflationary slow-roll parameter $\epsilon_{i} \equiv -H_{i}'/H_{i}$.
	
	The net curvature perturbation used in the body of this paper is the sum of the two components,
	\begin{align}
		\R = \R_{i}+\R_Q \, ,
	\end{align}
	and its power spectrum is
	\begin{align}
		\Delta_{\R\R}^2 = &\ \Delta^2_{\R_{i}} + \Delta^2_{\R_Q} \\
		= &\ \left(1 +\xi \right)  \Delta^2_{\R_Q} \, ,
	\end{align}
	where we have defined $\xi$ as the ratio of curvature perturbations generated by the inflaton, to curvature perturbations generated by $\delta Q$, 
	\begin{align}
		\label{eq:xi}
		\xi \equiv \frac{c_{Qi}P_{,Xi}}{2\epsilon_i\gamma^2}\frac{Q_i^2}{M_{pl}^2} \, .
	\end{align}
	The isocurvature perturbations used in this paper are 
	\begin{align}
		\I = &\ \frac{\delta \rho_Q}{\rho_Q} \\
	  	= &\ \frac{\partial\log \rho_Q}{\partial Q}\delta Q\lb\frac{c_Q^2-c_X^2}{c_Q^2-c_s^2}\rb \\
	  	\simeq &\  \frac{\partial\log \rho_Q}{\partial Q}\delta Q\lb\frac{c_Q^2-c_X^2}{c_Q^2+1}\rb \, ,
	\end{align}
	where $c_s^2$ is the synchronous gauge sound speed, defined in Eq.~(\ref{eq:cs}) and in the last line we have used the fact that outside of the horizon, $c_s^2\simeq-1$.
 
	The isocurvature auto- and cross-power spectra are then
	\begin{align}
		\Delta^2_{\R\I} = &\ \frac{Q_i}{\gamma}\frac{\partial \log \rho_Q}{\partial Q}\lb\frac{c_Q^2-c_X^2}{c_Q^2+1}\rb \Delta_{\R_Q}^2 \, , \\
		\Delta^2_{\I\I} = &\left(\frac{Q_i}{\gamma}\frac{\partial \log \rho_Q}{\partial Q} \lb\frac{c_Q^2-c_X^2}{c_Q^2+1}\rb \right)^2\Delta_{\R_Q}^2 \, .
	\end{align}
	The ratio of isocurvature to adiabatic perturbations is
	\begin{align}
		\frac{\Delta^2_{\I\I}}{\Delta^2_{\R\R}} = \frac{1}{(1+\xi)}\left( \frac{Q_i}{\gamma}\frac{\partial \log \rho_Q}{\partial Q} \lb\frac{c_Q^2-c_X^2}{c_Q^2+1}\rb\right)^2 \, ,
	\end{align}
	and the correlation between them is given by
	\begin{align}
		\frac{\Delta^2_{\R\I}}{\sqrt{\Delta^2_{\R\R}\Delta^2_{\I\I}}} = \frac{1}{\sqrt{1+\xi}} \, .
	\end{align}
	So, to have substantial correlations between $\R$ and $\I$, we need $\xi\lsim1$, and to have the isocurvature also be important ($\Delta_{\R\R}^2\sim \Delta_{\I\I}^2$) we need 
	\begin{align}
	\left(\frac{Q_i}{\gamma} \frac{\partial\log \rho_Q}{\partial Q}\lb\frac{c_Q^2-c_X^2}{c_Q^2+1}\rb\right)^{-2} \sim 1 \, .
	\end{align} 
	On the other hand, if we do not care about correlations between $\I$ and $\R$, $\xi$ can be large and we only need 
	\begin{align}
	\xi\left(\frac{Q_i}{\gamma}\frac{\partial\log \rho_Q}{\partial Q}\lb\frac{c_Q^2-c_X^2}{c_Q^2+1}\rb\right)^{-2} \sim 1  \, ,
	\end{align}
	to have isocurvature effects today. 

	From Eq.~(\ref{eq:xi}), we can see that both correlated and uncorrelated cases require 
	\begin{align}
		\frac{c_{Qi}P_{,Xi}}{2\epsilon_i}\left(M_{pl}\frac{\partial\log \rho_Q}{\partial Q}\right)^{-2}\lb\frac{c_Q^2+1}{c_Q^2-c_X^2}\rb^2\lsim 1 \, .
	\end{align}
	This can be rewritten in terms of $\Omega_Q$, and the slow-roll parameters, using Eq.~(\ref{eq:drho}) and the Friedmann equation, as
	\begin{align}
		\label{eq:condition}
		\frac{ c_{Qi} }{4}  \frac{ P_{,Xi} }{P_{,X}} \Omega_{Q} \lsim \epsilon_i \epsilon_{Q}\lb1+\frac{\eta_Q-2\epsilon_Q}{3(1+c_Q^2)}\rb^2 \, ,
	\end{align}
	where the subscript $i$ indicates that these quantities are evaluated during inflation, and quantities without this subscript are evaluated at late times. 

	For a standard kinetic term, Eq.~(\ref{eq:condition}) becomes
	\begin{align}
		\Omega_{Q} \lsim 4 \epsilon_i \epsilon_{Q}\lb1+\frac{\eta_Q-2\epsilon_Q}{6}\rb^2 \, ,
	\end{align}
	which cannot be satisfied by a field that dominates the energy density today and behaves as dark energy. 

	On the other hand, with a nonstandard kinetic term it becomes easier to satisfy Eq.~(\ref{eq:condition}). For the Lagrangian in Eq.~(\ref{eq:L}) we have 
	\begin{align}
		\label{eq:conditioncQ}
		c_{Q}\left(\frac{X_i}{X}\right)^{\frac{1-c_{Q}^2}{2c_Q^2}}\Omega_{Q} \lsim 4\epsilon_i \epsilon_{Q}\left(1+\frac{\eta_Q-2\epsilon_Q}{3(1+c_Q^2)}\right)^2 \, .
	\end{align}
	For the solution we considered in this paper, $\left({X_i}/{X}\right)^{\frac{1-c_{Q}^2}{2c_Q^2}} \sim H/H_i$, which allows the left-hand side of  Eq.~(\ref{eq:conditioncQ}) to be small even if, $\Omega_{Q}\sim 1$. The factor on the right-hand side involving $\eta_Q$ is given by the background solution,
	\begin{align}
		1+\frac{\eta_Q-2\epsilon_Q}{3(1+c_Q^2)} = -\frac{1}{3f} \, ,
	\end{align}
	where $f$ is defined in Eq.~(\ref{eq:f}). During matter domination, this gives $1+\eta_Q/(3(1+c_Q^2)) = \frac{3}{2}$. This factor, combined with other factors of $f$ from the expression for $X$ evaluated on the background solution, combine to a numerical constant of order $\sim10$ at most on the left-hand side of the above inequality.

	Finally, note that the tensors generated during inflation are
	\begin{align}
		\Delta^2_T = \frac{2 H^2_i}{\pi^2M_{pl}^2 }\, ,
	\end{align}
	so the tensor-to-scalar ratio is
	\begin{align}
	\frac{\Delta^2_T}{\Delta^2_{\R\R}} = \frac{16 \epsilon_i}{1+1/\xi}\, .
	\end{align}
	So, if $\xi \gg1$ (inflaton perturbations dominate) we have the usual expression for the tensor-to-scalar ratio, and as $\xi$ decreases, the tensor-to-scalar ratio gets even smaller. Note that even secondary production of gravitational waves by a spectator field like $Q$ cannot enhance the tensor-to-scalar ratio \cite{Fujita:2014oba}. 

\end{document}